\documentclass{raa}           

\usepackage{natbib}
\usepackage{longtable}
\usepackage{multirow}
\usepackage{graphicx,times}
\usepackage{natbib}
\usepackage{amssymb,amsmath}
\usepackage{siunitx}
\usepackage{booktabs}
\usepackage{threeparttable}
\usepackage{tabularx}
\usepackage{float}
\usepackage{lscape}
\bibpunct{(}{)}{;}{a}{}{,}

\newcommand{\CO}[3]{\ensuremath{{}^{#1}\mathrm{C}{}^{#2}{\mathrm{O}}}\ensuremath{#3}}

\newcommand{\HyMol}{\ensuremath{\mathrm{H_2}}}
\newcommand{\kps}{\ensuremath{\mathrm{~km~s^{-1}}}}
\newcommand{\kpc}{\ensuremath{~\mathrm{kpc}}}
\newcommand{\pc}{\ensuremath{~\mathrm{pc}}}
\newcommand{\Kelvin}{\ensuremath{~\mathrm{K}}}
\newcommand{\vlsr}{\ensuremath{V_{\mathrm{lsr}}}}
\newcommand{\Tex}{\ensuremath{T_{\mathrm{ex}}}}
\newcommand{\T}[2]{\ensuremath{T_{\mathrm{#1}}^{#2}}}
\newcommand{\cloudveldisp}{\ensuremath{\sigma_v}}

\newcommand{\cloudRsky}{\ensuremath{R_{\mathrm{eff}}}}
\newcommand{\cloudMass}{\ensuremath{M_{\mathrm{LTE}}}}

\newcommand{\cloudsurfacedensity}{\ensuremath{\mathit{\Sigma}}}
\newcommand{\losveldisp}{\ensuremath{{\sigma_v^{\mathrm{pix}}}}}

\newcommand{\pixRsky}{\ensuremath{R_{\mathrm{s}}}}
\newcommand{\virparam}{\ensuremath{\alpha_{\mathrm{vir}}}}
\newcommand{\Msun}{\ensuremath{~\mathrm{M}_{\odot}}}
\newcommand{\Msunpersqpc}{\ensuremath{~\mathrm{M}_{\odot}~\mathrm{pc^{-2}}}}

\newcommand{\pindex}{\ensuremath{\gamma}}
\newcommand{\pintercept}{\ensuremath{\sigma_{\mathrm{1pc}}}}

\usepackage[pagebackref=true, colorlinks=true, linkcolor=blue, citecolor=blue, filecolor=blue, urlcolor=blue]{hyperref}

\begin{document}

   \title{
      Revisiting the Velocity Dispersion-Size Relation in Molecular Cloud Structures 
    }

 \volnopage{ {\bf 20XX} Vol.\ {\bf X} No. {\bf XX}, 000--000}
   \setcounter{page}{1}

   \author{
    Haoran Feng \inst{1,2},
    Zhiwei Chen \inst{1},
    Zhibo Jiang \inst{1,3},
    Yuehui Ma \inst{1},
    Yang Yang \inst{4,1,2},
    Shuling Yu \inst{1,2},
    Dongqing Ge \inst{1,2},
    Wei Zhou\inst{1,2},
    Fujun Du \inst{1,2},
    Chen Wang \inst{5,1},
    Shiyu Zhang\inst{1,2},
    Yang Su \inst{1},
    and Ji Yang \inst{1}
   }
   \institute{ 
    Purple Mountain Observatory, Chinese Academy of Sciences, Nanjing 210023, China; {\it zwchen@pmo.ac.cn}\\
    \and
    University of Science and Technology of China, Chinese Academy of Sciences, Hefei 230026, China\\
    \and
    Center for Astronomy and Space Sciences, China Three Gorges University, Yichang 443000, China\\
    \and 
    Center for Astrophysics, Guangzhou University, Guangzhou 510006, China\\
    \and 
    National Astronomical Observatories, Chinese Academy of Sciences, Beijing 100101, China\\
\vs \no
   {\small Received 20XX Month Day; accepted 20XX Month Day}
}

\abstract{
Structures in molecular ISM are observed to follow a power-law relation between the velocity dispersion and spatial size, known as Larson's first relation, which is often attributed to the turbulent nature of molecular ISM and imprints the dynamics of molecular cloud structures.
Using the $\element[][13]{CO}~(J=1-0)$ data from the Milky Way Imaging Scroll Painting survey, we built a sample with 360 structures having relatively accurate distances obtained from either the reddened background stars with {\it Gaia} parallaxes or associated maser parallaxes, spanning from $0.4$ to $\sim 15\kpc$.
Using this sample and about 0.3 million pixels, we analyzed the correlations between velocity dispersion, surface/column density, and spatial scales. 
Our structure-wise results show power-law indices smaller than 0.5 in both the $\cloudveldisp$-$\cloudRsky$ and $\cloudveldisp$-$\cloudRsky \cdot \cloudsurfacedensity$ relations. 
In the pixel-wise results, the $\losveldisp$ is statistically scaling with the beam physical size ($\pixRsky\equiv \Theta D/2$) in form of $\losveldisp \propto \pixRsky^{0.43 \pm 0.03}$.  
Meanwhile, $\losveldisp$ in the inner Galaxy is statistically larger than the outer side.
We also analyzed correlations between $\losveldisp$ and the $\HyMol$ column density $N(\HyMol)$, finding that $\losveldisp$ stops increasing with $N(\HyMol)$ after $\gtrsim 10^{22}~{\mathrm{cm^{-2}}}$. 
The structures with and without high-column-density ($> 10^{22}~\mathrm{cm^{-2}}$) pixels show different $\losveldisp \propto N(\HyMol)^{\xi}$ relations, where the mean (std) $\xi$ values are $0.38~(0.14)$ and $0.62~(0.27)$, respectively.
}
   \authorrunning{H. Feng et al.}            
   \titlerunning{Revisiting the Velocity Dispersion-Size Relation}  
   \maketitle

%
\section{Introduction} \label{sect:intro}
Molecular clouds (MCs) are essential to the material cycle and evolution of galaxies since they are the birthplaces of stars \citep[e.g.,][]{2007ARA&A..45..339B,2020SSRv..216...76B}.  
Although interstellar molecular Hydrogen ($\HyMol$) constitutes most of the gas in MCs, it is the carbon monoxide (\CO{}{}{}) and its isotopic molecules that are most used as the tracers of molecular gas \citep{2015ARA&A..53..583H}.
From the observational perspective, the MCs are more appropriately defined by the connected structures in position-position-velocity (PPV) space of the most extended \CO{12}{}{~(J=1-0)} emission \citep[e.g.,][]{1987ApJ...319..730S,2022AJ....164...55Y}.
The rare CO isotopologues, for example, \CO{13}{}{} and \CO{}{18}{}, can help trace the substructures of MCs.
The large line widths have been identified since the first observation of \CO{}{}{~(J=1-0)} by \citet{1970ApJ...161L..43W}.
In consideration of the cold ($T \sim 10~\Kelvin$) and dense ($n \gtrsim 10^{2}~\mathrm{cm^{-3}}$) environment in MCs, the high-velocity dispersion indicates supersonic (Mach number $\mathcal{M} \gg 1$) motions that are often attributed to supersonic turbulence \citep{2013MNRAS.436.3247K}.
Meanwhile, the dynamic properties of MCs are fundamental for understanding cloud formation, dissipation, and star-forming activities, making the MC velocity analysis of vital importance.
Insights about MCs' gravity, turbulence, and magnetic fields can be obtained by analyzing the velocity structures.
The techniques applied by previous studies include but are not limited to: velocity centroids analysis \citep[VCA; e.g.,][]{1984ApJ...277..556S, 2020ApJ...901...11H}, $\Delta$-variance \citep[e.g.,][]{2001A&A...366..636B}, one-point PDF \citep[e.g.,][]{2000ApJ...535..869K}, principal component analysis \citep[PCA; e.g.,][]{1997ApJ...475..173H, 2023MNRAS.519.5427D}, and velocity gradient techniques for magnetic field \citep[e.g.,][]{2017ApJ...835...41G, 2022ApJ...934...45Z}, etc.
These methods and techniques have fully utilized observational or simulated data and enhanced our understanding of MC dynamics.

Before this prosperity, \citet{1981MNRAS.194..809L} has concluded three basic relations of MCs:
(1) The velocity dispersion - size relation, $\cloudveldisp \propto R^{\pindex}$, $\pindex=0.38$;
(2) MCs are gravitationally bound or in virial equilibrium (VE), with the virial parameter $\alpha_{\mathrm{vir}}=E_{k} / E_{G}\sim 1$;
(3) The volumetric density - size relation $n \propto R^{-1.1}$.
These relations are not independent, as any two of them imply the other. 
The third one is equivalent to a mass-size relation of $M \propto R^{2}$, which indicates a constant surface density of molecular clouds.
However, \citet{2022RAA....22g5006X} found that the power-law index of the mass-size relation is significantly affected by the column density threshold, with which the boundaries of molecular cloud structures are defined. 
Besides, molecular tracers that are effective in different ranges of densities may lead to different mass-size scaling relations. 
This work will use a single molecular tracer and concentrate on the first relation.

Larson's first relation correlates the global velocity dispersion with the size of MCs or the substructures. 
The global velocity dispersion of a single MC is analog to the structure function at the cloud scale \citep[e.g.,][]{2004ApJ...615L..45H}.
The power-law index $\pindex=0.38$ has been interpreted as incompressible turbulence in high Reynold number fluid, as it is close to the value of $1/3$ predicted by the theory of \citet{1941DoSSR..30..301K}.
However, it is conceivable that the interstellar gas is compressible, and the power-law index is expected to be larger than $1/3$ \citep{2021ApJ...906L...4C}.
Due to its importance in the dynamics of molecular ISM, the relation has been continuously revisited \citep[e.g.,][]{1987ApJ...319..730S,1992ApJ...384..523F,1998ApJ...504..223G,2023ApJ...949...46H} and questioned \citep[e.g.,][]{2018MNRAS.477.2220T}.
The current most widely accepted power-law index for the relation is 0.5 \citep{1987ApJ...319..730S}, but other values also exist, for example, 0.21 \citep{1995ApJ...446..665C}, 0.56 \citep{2004ApJ...615L..45H}, 0.70 \citep{2022MNRAS.513..638Z}.
These values are obtained through different tracers among different kinds of objects, making it hard to put them in the same context of discussion. 

An alternative approach other than cloud structure catalogs is the line of sight analysis \citep{2016ApJ...831...16L,2020ApJ...901L...8S,2021MNRAS.502.1218R}.
It measures the gas properties along the lines of sight at one or more fixed spatial scales. 
In this manner, one can reduce the difference caused by various structure identification methods.
Inspired by these extragalactic works, we attempt to conduct similar pixel-by-pixel analyses on the molecular gas in the Milky Way (MW).
However, determining distances to every pixel is impractical.
We still cluster the molecular line emission into individual structures to find reliable distance measurements.

In this work, we analyze the correlations between velocity dispersion and other physical properties of 360 \CO{13}{}{} structures and the pixels therein. 
The paper is organized as follows: 
Section \ref{sect:data} describes the CO data and distance measurements. 
The structure identification and derivation of the physical properties are presented in Sect.~\ref{sect:methods}.
In Sect. \ref{sect:structures} and \ref{sect:pixelwiseresults}, we present the structure-wise and pixel-wise results, respectively.
We discuss the Keto-Heyer diagram, virial parameter, and the spatial scale of the measured velocity dispersion in Sect.~\ref{sect:discuss}.
Finally, a summary is provided in Sect.~\ref{sect:summary}.


\section{Data} \label{sect:data}

\subsection{MWISP CO Data}  \label{sect:mwisp_data}

In this work, we extract \CO{13}{}{} structures from the data cubes of the Milky Way Imaging Scroll Painting (MWISP) survey to explore the velocity dispersion-size relation.
\citet{2019ApJS..240....9S} have given a detailed description of the MWISP survey, and a preliminary noise analysis has been performed by \citet{2021RAA....21..304C}.
We summarize the major characteristics of the MWISP \CO{}{}{} data here: 
(a) The observations are taken in position-switch On-The-Fly (OTF) mode with half power beam widths (HPBWs) of $\sim49\arcsec$ for \CO{12}{}{~(J=1-0)}, $\sim 52\arcsec$ for \CO{13}{}{} and \CO{}{18}{~(J=1-0)}. 
(b) The data are gridded into $30'' \times 30''$ pixels.
(c) At channel widths of $0.16, 0.17$ and $0.17\kps$, the typical rms noise levels, $\sigma_{\rm rms}$, are $\sim 0.5, \sim0.3$ and $\sim0.3 \Kelvin$ for \CO{12}{}{}, \CO{13}{}{}, and \CO{}{18}{}, respectively.

We use the MWISP \CO{12}{}{} data to derive excitation temperature $(\T{ex}{})$ in calculating the column densities under the assumption of local thermodynamic equilibrium (LTE).
The line width of the optically thick \CO{12}{}{} could be heavily influenced by the opacity broadening effect \citep{2016A&A...591A.104H}.
Therefore, the velocity dispersion and column density are derived from the \CO{13}{}{} data.
In our case, most lines of sight have \CO{13}{}{} central optical depth $\tau_{\CO{13}{}{}} < 1.0$, 
indicating that we can safely ignore the opacity broadening effect. 

\subsection{Distance Measurements}  \label{sect:distance}

The spatial size of a structure is directly proportional to its distance from us.
The uncertainty of the velocity dispersion-size relation is dominated by the error of distances.
Many previous works utilized kinematic distances, which heavily rely on models and suffer from near/far ambiguity toward the inner galaxy. 
Therefore, to have an accurate estimation of the spatial scale, we employ recent distance measurement results with maser parallaxes \citep{2019ApJ...885..131R,2020PASJ...72...50V,2021ApJS..253....1X,2022AJ....163...54B, 2022PASJ...74..209S,2022ApJS..262...42L, 2024AJ....167..267B} and \textit{Gaia} \citep{2016A&A...595A...1G} parallaxes of reddened background stars \citep{2021ApJ...922....8Y,2024A&A...685A..39M,2024AJ....167..220Z,2024ApJ...966..202Z}.
These measurements also provided the radial velocities (\vlsr) that help match our structures.

\section{Methods} \label{sect:methods}

\subsection{Structure Identification} \label{sect:find_structs}

We used the newly developed ISMGCC (InterStellar Medium Gaussian Component Clustering) method \citep{2024arXiv240901181F} to acquire the structures for our analysis. 
This method uses the Gaussian decomposition results of data cubes as the input and returns the clustering results where each structure is a collection of Gaussian components (GCs).
It could give rational structure segmentation in crowded regions. 
Meanwhile, it is sensitive to weak emissions and could retain most of the flux.
Most importantly, the ISMGCC method is designed to distinguish multiple velocity components along the line of sight, without which the velocity dispersion might be overestimated because blended line profiles from multiple structures could significantly increase the measured velocity dispersion.

\begin{figure}[t!]
   \centering
   \includegraphics[width=\textwidth]{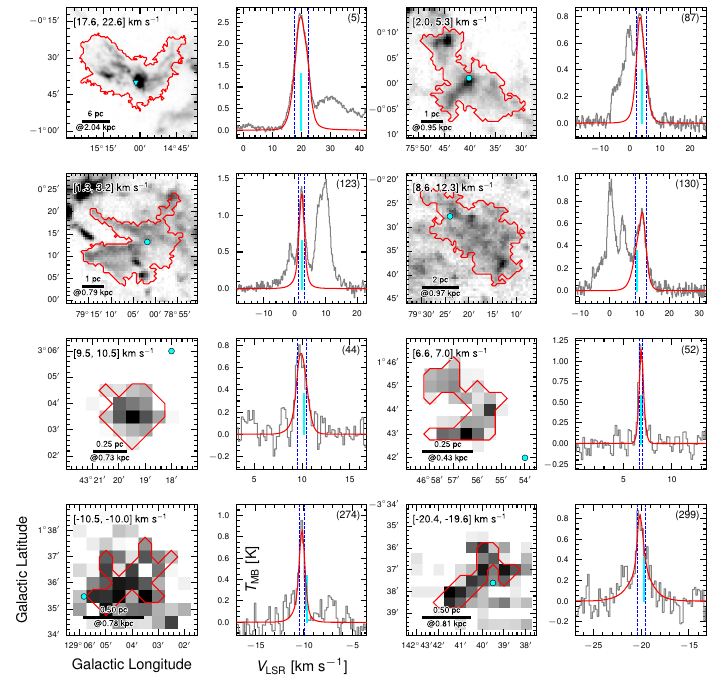}
   \caption{
   Eight examples of the structures identified with ISMGCC.
   The two rows on top contain four structures in the crowded regions. 
   The other two rows at the bottom contain the four structures with the lowest $\cloudMass$. 
   The number in parentheses on the top right corner of each spectral panel denotes the structure ID.
   The image on the left side of each spectral panel is an integrated intensity map within the velocity interval indicated by the blue dashed lines in the spectral panel.
   These velocity intervals are defined by the half-peak intensities of the recovered average spectra from the Gaussian components (GCs) clustered by ISMGCC.
   The red curves show the recovered average spectra, while the gray profiles are the raw average spectra.
   Both kinds of average spectra are within the spatial boundary denoted by the red contour in each image panel. 
   Locations of the distance measurements are marked as cyan triangles (for maser parallaxes) and dots (for other methods).
   The $\vlsr$ of the distance measurements are marked by the cyan vertical line in the spectral panels.
   }
   \label{fig:examples}
\end{figure}

We employed \textsc{GaussPy+} \citep{2019A&A...628A..78R} with the default parameters to decompose the MWISP \CO{13}{}{} data cubes around the $(l, b, \vlsr)$ coordinates with distance measurements, during which the values of two smoothing parameters $\alpha_1=2.18$ and $\alpha_2=4.94$ were taken from the tests on MWISP \CO{13}{}{} by \citet{2020A&A...633A..14R}.
Then the ISMGCC method processed the output table of \textsc{GaussPy+} with the optimized parameters in \citet{2024arXiv240901181F} to find the structures. 
The identified structures are based on \CO{13}{}{} emissions and have no specific geometries.  
Because the most suitable emission line to define molecular clouds is \CO{12}{}{~(J=1-0)} \citep[e.g.,][]{1987ApJ...319..730S,2022AJ....164...55Y}, the most suitable term to describe the sample analyzed in this work should be ``\CO{13}{}{} structures''. 
Figure~\ref{fig:examples} shows some examples of the identified \CO{13}{}{} structures. 
These examples demonstrate two features of the ISMGCC method. 
Firstly, it could distinguish structures toward the crowded regions, as shown by the four examples on top;
Secondly, it is sensitive to structures with weak emission, e.g., the four examples at the bottom, because of no global signal-to-noise ratio cutoff on the input data cube. 

A structure is considered associated with a given distance measurement $(l, b, \vlsr, D)$ point if the following two conditions are satisfied:
\begin{enumerate}
   \item At least one pixel in the structure has angular separation less than $2'$ from $(l, b)$;
   \item The velocity difference between $\vlsr$ and centroid velocity of the structure's spectrum on the nearest pixel is less than the spectrum's FWHM.
\end{enumerate}
When a structure is associated with multiple distances, we only retained the one with the least velocity difference. 
The total number of structures with associated distance measurements is 431.
Besides, the minimal pixel number for ISMGCC to consider a structure valid was set to 16, about four beam sizes of the telescope. 
To improve the reliability of the analysis results, we retained only the structures having great consistency between the raw average spectrum (gray profiles in the spectral panels of Fig.~\ref{fig:examples}) and the recovered average spectrum (red curves).
A recovered average spectrum should recover at least 85\% flux of the raw average spectrum in the velocity interval demonstrated by The blue dashed lines in Fig.~\ref{fig:examples}.
The half-max value of the recovered average spectrum defines this interval. 
Besides, structures with incomplete boundaries were also removed. 
The final number of structures in our analysis is 360. 

\begin{figure}[h!]
   \centering
   \includegraphics[width=0.9\textwidth]{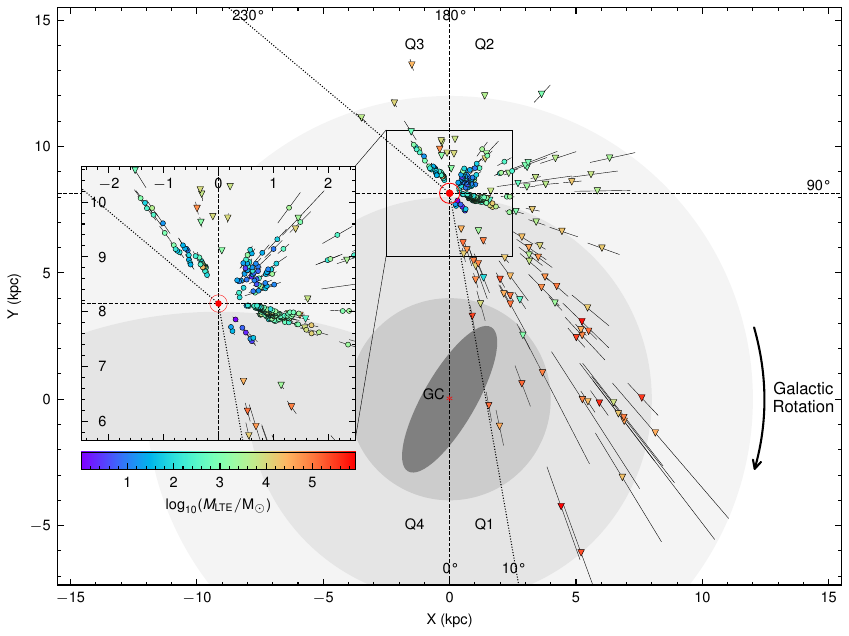}
   \caption{
   Spatial distribution of the \CO{13}{}{} structures on the face-on view of the galaxy from the north Galactic pole. 
   The colored inverse triangles and circles mark the \CO{13}{}{} structures associated with maser parallaxes and distance measurements with background star parallax, respectively.
   The Sun ($\odot$) is located at $(0, 8.15)\kpc$ in the galactocentric frame, around which a zoom-in view within $2.5\kpc$ is shown.
   Shaded concentric circles around the Galactic Center (GC) have radii of $4, 8$, and $12\kpc$, which are the same as figure 1 of \citet{2014ApJ...783..130R}, while the ellipse is a schematic representation of the Galactic bar. 
   }
   \label{fig:topview}
\end{figure}

Figure \ref{fig:topview} shows the plan view of the Galactic distribution of these structures.
The inverse triangles and circles denote the \CO{13}{}{} structures associated with the maser parallaxes \citep{2019ApJ...885..131R,2020PASJ...72...50V,2021ApJS..253....1X,2022AJ....163...54B, 2022PASJ...74..209S,2022ApJS..262...42L, 2024AJ....167..267B} and reddened background stars \citep{2021ApJ...922....8Y,2024A&A...685A..39M,2024AJ....167..220Z,2024ApJ...966..202Z}, respectively.
The 95 structures associated with masers are widely distributed throughout the galaxy at various distances, while those with background stars are relatively concentrated in the solar neighborhood.

\subsection{Derivation of Physical Properties} \label{sect:property}

The physical properties involved in our analysis include the size, velocity dispersion, column density/surface density, and mass of the structures. 
The first three properties exist at two levels, pixels and structures.
Therefore, for each of them, we will first introduce its derivation among pixels, then structures. 

For the physical size, we define the radius of a single pixel on the plane of the sky as 
\begin{equation}
    \pixRsky \equiv \frac{1}{2} \Theta D
\end{equation}
in which $\Theta \approx 52''$ is the half power beam width of the telescope at $\sim 110 \mathrm{~GHz}$.
$D$ is the distance of the parent structure. 
The effective radius of a structure is calculated by the following equation \citep{1994ApJ...433..117L}
\begin{equation}
    \cloudRsky = \frac{1}{2}D  \sqrt{\frac{4A}{\pi} - \Theta^2},
\end{equation}
where $A$ is the angular area.

The velocity dispersion for each pixel is denoted as $\losveldisp$,  while $\cloudveldisp$ is reserved for the structures. 
As the structures identified by ISMGCC are collections of Gaussian components (GCs), one can generate a noise-free data cube for each structure with its GCs. 
This recovered data cube can derive $\losveldisp$ and $\cloudveldisp$.
We clip the recovered cube at $2\sigma_{\mathrm{rms}}$ level and use the square root of its Moment 2 map to calculate $\losveldisp$.
It can also be utilized to create a noise-free average spectrum of the structure, also known as the recovered average spectrum in Sect.~\ref{sect:property}.
The red curves in the spectral panels of Fig.~\ref{fig:examples} are examples of such recovered average spectra, where the second moment is applied to calculate $\cloudveldisp$. 

We use the line pair of \CO{12}{}{} and \CO{13}{}{~J=(1-0)} to derive the \CO{13}{}{} column density $N(\CO{13}{}{})$ for each pixel with the assumption of Local Thermodynamic Equilibrium (LTE).
To derive the excitation temperature $\T{ex}{}$ on each pixel, we extract the \CO{12}{}{} peak intensity (\T{peak}{\CO{12}{}{}}) within the velocity interval defined by the half-max value of \CO{13}{}{} spectrum in the recovered data cube.
Within this velocity interval we can also extract \T{peak}{\CO{13}{}{}} to calculate the central optical depth $\tau_{13}$ and use it to make correction on $N(\CO{13}{}{})$.
Detailed formulas are given in Appendix~\ref{ap:columndensity}.
Following \citet{2009ApJ...699.1092H}, we convert $N(\CO{13}{}{})$ to $N(\HyMol)$ with the constant abundance ratio $[\HyMol/\CO{12}{}{}]=1.1\times 10^{4}$ \citep{1982ApJ...262..590F} and the one with the gradient $[{}^{12}\mathrm{C}/{}^{13}\mathrm{C}]=6.2 (R_{\mathrm{gal}}/ 1 \kpc) + 18.7$ \citep{2005ApJ...634.1126M}, where $R_{\mathrm{gal}}$ is the galactocentric radius.

The mass surface density of an entire structure is calculated by
\begin{equation}
\cloudsurfacedensity =\mu_{\HyMol} m_{\mathrm{H}}\langle N({\HyMol})\rangle 
\end{equation}
where $\mu_{\HyMol}=2.8$ is the molecular mass per hydrogen molecule \citep{2008A&A...487..993K},
$m_{\mathrm{H}}$ is the H-atom mass,
and $\langle N({\HyMol}) \rangle$ is the average $\HyMol$ column density derived on its belonging pixels.  
The total LTE mass of a cloud structure is
\begin{equation}
    \cloudMass =   \pi\cloudRsky^2 \cdot \cloudsurfacedensity.
\end{equation}

The surface density $\cloudsurfacedensity$ is expected to be dominated by the low-density regions of the identified structures, which is often close to the lower detection limit. 
\citet{2009ApJ...699.1092H} defined another area within the half-max isophote of the peak column density in each structure to increase the dynamic range of $\cloudsurfacedensity$.
We reproduced this procedure and calculated the physical parameters described above.
The physical properties of the 360 structures and their half-max isophote subregions are listed in Table \ref{table:properties}.
In the next sections, we will show their distributions and use them to analyze the velocity dispersion-size relation.
These structures contain a total of 366,605 pixels. 
In Sect.~\ref{sect:pixelwiseresults}, we will use the pixels with peak signal-to-noise ratios greater than five to analyze the distributions and relations between $\pixRsky, N(\HyMol)$, and $\losveldisp$.
The number of pixels in the $5\sigma_{\mathrm{rms}}$ subset is 289,696, about $80\%$ of the total.

\begin{landscape}
\begin{table}
\begin{threeparttable}
\caption{
    Properties of the Identified \CO{13}{}{} Structures
}\label{table:properties}
\begin{tabular}{
    @{}
    c 
    SS[retain-explicit-plus]
    S
    SScc
    SScc 
    rlc  
    @{}
    }
\hline\hline   \\
  & & & &
    \multicolumn{4}{c}{Within the Entire Structure} 
  & \multicolumn{4}{c}{Within Half-Max Isophote of $N(\HyMol)$} 
  & \multicolumn{3}{c}{Distance Information}\\
  \cmidrule(lr){5-8} \cmidrule(lr){9-12} \cmidrule(lr){13-15}
  No. & $l$            & $b$            & $\vlsr$                & $\cloudveldisp$        & \multicolumn{1}{c}{$\cloudRsky$  }  & \multicolumn{1}{c}{\cloudsurfacedensity} & $\cloudMass$  & \cloudveldisp         & \cloudRsky                        & \cloudsurfacedensity & \cloudMass & \multicolumn{1}{c}{$D$} & \multicolumn{1}{c}{Name} & Ref. \\
      & $\mathrm{deg}$ & $\mathrm{deg}$ & $\mathrm{km\,s^{-1}}$  & $\mathrm{km\,s^{-1}}$  & \multicolumn{1}{c}{$\mathrm{pc}$ }  & \multicolumn{1}{c}{$\Msunpersqpc$      } & $\Msun$       & $\mathrm{km\,s^{-1}}$ & \multicolumn{1}{c}{$\mathrm{pc}$} &    $\Msunpersqpc$    & $\Msun$    & \multicolumn{1}{c}{$\mathrm{pc}$}  &  &   \\
\hline
1 & 10.62 & -0.38 & -2.4 & 2.89 & 24.8 & 1.3e+02 & 2.6e+05 & 3.47 & 2.0 & 7.4e+02 & 9.0e+03 & ${4.95}^{+0.51}_{-0.43}$ & G010.62$-$00.38* & 1 \\
2 & 10.47 & +0.03 & 68.2 & 3.38 & 19.6 & 1.1e+02 & 1.3e+05 & 2.50 & 0.8 & 2.5e+03 & 5.0e+03 & ${8.55}^{+0.63}_{-0.55}$ & G010.47$+$00.02* & 1 \\
5 & 15.02 & -0.67 & 20.1 & 2.94 & 8.9 & 4.6e+02 & 1.1e+05 & 2.33 & 0.7 & 8.4e+03 & 1.4e+04 & ${2.04}^{+0.19}_{-0.16}$ & M17* & 2 \\
10 & 19.37 & -0.02 & 25.9 & 2.11 & 12.8 & 1.6e+02 & 8.3e+04 & 1.70 & 1.7 & 6.2e+02 & 5.4e+03 & ${2.84}^{+0.69}_{-0.47}$ & G019.36$-$00.03* & 7 \\
11 & 19.62 & -0.26 & 40.6 & 4.22 & 46.0 & 1.3e+02 & 8.6e+05 & 2.16 & 0.7 & 2.7e+02 & 4.8e+02 & ${13.16}^{+2.23}_{-1.66}$ & G019.60$-$0.23* & 3 \\
13 & 20.08 & -0.13 & 42.5 & 4.12 & 25.1 & 1.1e+02 & 2.2e+05 & 2.06 & 2.4 & 6.7e+02 & 1.3e+04 & ${15.15}^{+2.71}_{-1.99}$ & G020.08$-$0.13* & 3 \\
15 & 21.77 & -0.12 & 63.2 & 2.02 & 1.1 & 2.4e+01 & 9.3e+01 & 2.07 & 0.7 & 3.3e+01 & 4.3e+01 & ${3.61}^{+0.63}_{-0.47}$ & IRAS~18286$-$0959* & 2 \\
16 & 22.04 & +0.22 & 51.9 & 3.41 & 16.5 & 9.5e+01 & 8.2e+04 & 1.79 & 1.7 & 4.4e+02 & 3.8e+03 & ${3.01}^{+0.59}_{-0.42}$ & G022.03$+$00.22* & 7 \\
17 & 26.92 & -3.56 & 16.3 & 0.43 & 0.7 & 2.6e+01 & 3.4e+01 & 0.41 & 0.4 & 4.2e+01 & 1.9e+01 & ${0.56}^{+0.08}_{-0.08}$ & G026.9$-$03.5 & 8 \\
18 & 27.23 & +0.13 & 113.1 & 1.04 & 3.4 & 3.3e+01 & 1.2e+03 & 0.82 & 0.7 & 8.8e+01 & 1.4e+02 & ${6.33}^{+0.62}_{-0.52}$ & G027.22$+$0.14* & 5 \\
20 & 28.07 & -2.12 & 18.8 & 0.65 & 0.6 & 1.7e+01 & 1.8e+01 & 0.67 & 0.3 & 2.8e+01 & 6.5e+00 & ${0.48}^{+0.03}_{-0.03}$ & G027.8$-$02.1 & 4 \\
28 & 33.64 & -0.23 & 60.7 & 1.21 & 11.0 & 6.2e+01 & 2.3e+04 & 1.24 & 2.3 & 3.0e+02 & 4.9e+03 & ${9.90}^{+0.50}_{-0.50}$ & G033.64$-$00.22* & 11 \\
30 & 35.50 & -0.03 & 54.7 & 3.77 & 24.8 & 2.2e+02 & 4.3e+05 & 3.37 & 2.6 & 8.4e+02 & 1.8e+04 & ${10.20}^{+0.60}_{-0.60}$ & G035.57$-$00.03* & 11 \\
40 & 41.51 & +2.27 & 17.9 & 1.04 & 0.3 & 1.1e+01 & 3.1e+00 & 0.82 & 0.2 & 1.5e+01 & 1.3e+00 & ${0.92}^{+0.10}_{-0.10}$ & G041.5$+$02.3 & 8 \\
48 & 44.57 & +2.74 & 14.7 & 0.43 & 0.4 & 1.5e+01 & 6.0e+00 & 0.35 & 0.1 & 2.5e+01 & 1.6e+00 & ${0.62}^{+0.05}_{-0.05}$ & G044.5$+$02.7 & 4 \\
56 & 50.28 & -0.39 & 14.7 & 1.41 & 9.1 & 6.3e+01 & 1.7e+04 & 1.62 & 1.3 & 2.2e+02 & 1.1e+03 & ${7.10}^{+1.10}_{-0.80}$ & G050.28$-$00.39* & 6 \\
65 & 72.37 & +2.11 & -2.2 & 1.36 & 5.2 & 1.5e+02 & 1.2e+04 & 1.23 & 0.7 & 4.2e+02 & 6.4e+02 & ${1.78}^{+0.12}_{-0.11}$ & ID: 1 & 9 \\
66 & 72.42 & +0.37 & 4.5 & 1.07 & 3.8 & 5.0e+01 & 2.2e+03 & 0.93 & 0.5 & 1.7e+02 & 1.5e+02 & ${2.48}^{+0.20}_{-0.17}$ & ID: 2 & 9 \\
319 & 198.97 & +1.43 & 7.4 & 0.89 & 2.9 & 3.9e+01 & 1.1e+03 & 0.94 & 0.5 & 1.4e+02 & 1.1e+02 & ${0.71}^{+0.05}_{-0.05}$ & ID: 6 & 10 \\
320 & 199.37 & +1.05 & 5.1 & 0.58 & 1.0 & 1.6e+01 & 5.1e+01 & 0.47 & 0.6 & 2.2e+01 & 2.5e+01 & ${0.70}^{+0.08}_{-0.09}$ & ID: 23 & 10 \\
\hline
\end{tabular}
    \begin{tablenotes}
      \small
      \item (a) The distance references in the last column are: (1) \citet{2019ApJ...885..131R}, (2) \citet{2020PASJ...72...50V}, (3) \citet{2021ApJS..253....1X}, (4) \citet{2021ApJ...922....8Y}, (5) \citet{2022AJ....163...54B}, (6) \citet{2022PASJ...74..209S}, (7) \citet{2022ApJS..262...42L}, (8) \citet{2024A&A...685A..39M}, (9) \citet{2024AJ....167..220Z}, (10) \citet{2024ApJ...966..202Z}, and (11) \citet{2024AJ....167..267B};
      \item (b) The asterisk (*) suffix in the distance source name denotes it is a maser parallax result.
      \item (c) This table is available in its entirety online (\url{https://doi.org/10.57760/sciencedb.14853}). Only the first one or two records associated with each distance reference are shown here.
    \end{tablenotes}
\end{threeparttable}
\end{table}
\end{landscape}

\section{Structure-wise Results}  \label{sect:structures}

In this section, we first present the physical property distributions of the \CO{13}{}{} structures, then demonstrate the $\cloudveldisp$-$\cloudRsky$ relation. 
The impact of surface density on this relation is also visited.

\subsection{Physical Property Distributions} \label{sect:structure-properties}

\begin{figure}[h!]
   \centering   
   \includegraphics[width=\textwidth]{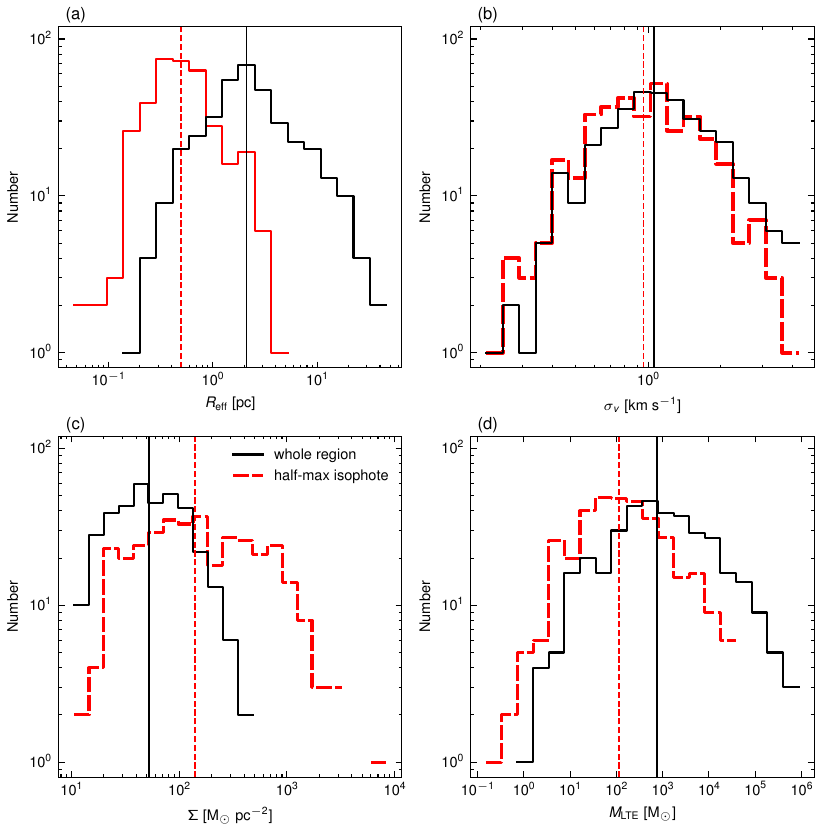}
   \caption{Physical property distributions of the identified structures. 
   The solid black and dashed red histograms correspond to the values within the entire structures and the $N(\HyMol)$ half-max isophotes, respectively. 
   The vertical lines across each panel denote the median values.
   }
   \label{fig:hist-structs}
\end{figure}

Here we describe the $\cloudRsky, \cloudveldisp, \cloudsurfacedensity, $ and $\cloudMass$ distributions of the \CO{13}{}{} structures and the subregions defined by their $N(\HyMol)$ half-max isophotes. 
The histograms are shown in Fig.~\ref{fig:hist-structs}.
The sample number of the whole regions and half-max isophotes are 360 and 352, respectively, because eight half-max isophotes contain less than three pixels leading to invalid $\cloudRsky$ values.
The $\cloudRsky$ of the whole regions are in the range of $[0.16, 46]~\mathrm{pc}$, spanning two orders of magnitude with a median value of $2.1~\mathrm{pc}$, similar to the common size of molecular clouds. 
The largest structures could be around tens of parsecs, corresponding to the Giant Molecular Cloud (GMC) or GMC Complex. 
The most massive structure has $\cloudMass \gtrsim 8\times 10^{5} \Msun$, while the median value is $764~\Msun$.

In contrast to the large dynamic range of $\cloudMass$, the $\cloudsurfacedensity$ values of the whole regions are limited in a small interval of $[10.6, 460]~\Msunpersqpc$, because the projected area of a resolved molecular cloud is commonly dominated by the pixels with the lowest detectable column densities.
\citet{2009ApJ...699.1092H} used the subregions within the $N(\HyMol)$ half-max isophote to increase the $\cloudsurfacedensity$ dynamic range. 
We follow their work and show the half-max isophote distributions as the red histograms in Fig.~\ref{fig:hist-structs}.
The value span of $\cloudsurfacedensity$ is instantly increased from $[10.6, 460]~\Msunpersqpc$ of the whole regions to $[11.6, 8.4\times 10^{3}] \Msunpersqpc $, as shown in Fig.~\ref{fig:hist-structs}(c).
The median value is also increased from $52.7\Msunpersqpc$ to $141\Msunpersqpc$.
Meanwhile, shrinking the spatial boundary would naturally decrease $\cloudRsky$ and $\cloudMass$. 
The value ranges of $\cloudRsky$ and $\cloudMass$ in the half-max isophotes are $[0.05, 4.2]~\mathrm{pc}$ and $[0.15, 3.0\times 10^{4}]\Msun$, respectively, while the median values decrease to $0.5~\mathrm{pc}$ and $113 \Msun$.
The size and mass distributions of the half-max isophotes are similar to the common definition of molecular clumps, as molecular clumps often refer to molecular structures with sizes of $\sim 1 \pc$ and masses from $\sim10$ to $\sim10^{3}\Msun$ \citep[e.g.,][]{2011piim.book.....D, 2013ApJ...779..185K}.

Shrinking the spatial boundary does not significantly impact the $\cloudveldisp$ distribution. 
As shown in Fig.~\ref{fig:hist-structs}, the median values of $\cloudveldisp$ in the whole regions and half-max isophotes are $1.06$ and $0.96\kps$, respectively. 
Meanwhile, their lower and upper limits are similar, specifically, $[0.23, 4.21] \kps$  for the whole regions and $[0.21, 3.61] \kps$ for the half-max isophotes.
This could impact the $\cloudveldisp$-$\cloudRsky$ relation when combing the whole region and half-max isophote samples, as they have different $\cloudRsky$ distributions.
We will explicitly show this in the next subsection.

\subsection{The $\cloudveldisp$-$\cloudRsky$ Relation} \label{sect:larson1}

Here we examine the structure-wise scaling relation between $\cloudRsky$ and $\cloudveldisp$ in the form of 
\begin{equation} \label{eq:firstlarsoncloud}
    \frac{\cloudveldisp}{\pintercept} = \left(\frac{\cloudRsky}{1~\mathrm{pc}}\right)^{\pindex},
\end{equation}
where $\pindex$ is the power-law index, and $\pintercept$ is the scaling coefficient, the velocity dispersion at $1 \pc$ scale.
This relation is known as Larson's first relation \citep{1981MNRAS.194..809L}, and has been commonly considered as the result of interstellar turbulence, while the $\pindex$ value and interpretations are still under debate. 
Using the sample of 360 structures extracted from the MWISP \CO{13}{}{} data, this multi-structure, single-tracer exploration should be classified as the Type 2 relation according to \citet{1998ApJ...504..223G}.

\begin{figure}[h!]
   \centering
   \includegraphics[width=\textwidth]{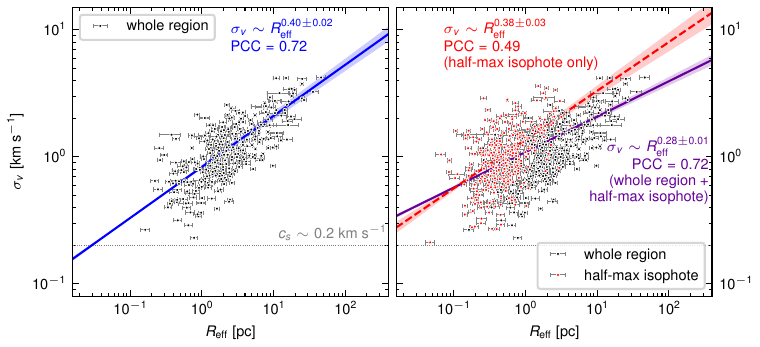}
   \caption{
      Relation between the velocity dispersions ($\cloudveldisp$) and effective radii ($\cloudRsky$) of the \CO{13}{}{} structures and their $N(\HyMol)$ half-max isophotes. 
      The left panel contains only the whole regions and both kind of regions are shown in the right panel. 
      The straight lines with error bands are the ODR fitting results of the power-law function. 
      Alongside each power-law fitting result is the Pearson Correlation Coefficient (PCC).
      The horizontal line with $c_s \sim 0.2 \kps$ denotes the typical sonic speed in molecular clouds. 
   }

   \label{fig:larson1}
\end{figure}

In Fig. \ref{fig:larson1}, we plot $\cloudveldisp$ of the structures against their effective radii (\cloudRsky).
To evaluate the impact of adding the half-max isophote samples on the relation, we use two panels in Fig.~\ref{fig:larson1} to draw the data points and fitting results, where the left panel only includes the whole regions and the right panel contains two configurations, half-max ishophtes only and the union of the whole regions and half-max isophotes.
Following \citet{2016ApJ...822...52R}, we use Orthogonal Distance Regression \citep[ODR;][]{2020SciPy-NMeth} to fit Eq. \ref{eq:firstlarsoncloud} on these data points.  
The ODR can handle measurement errors in the explanatory variable, i.e., $\cloudRsky$. 
The blue, red, and violet straight lines in Fig.~\ref{fig:larson1} represent the power-law fitting results on sample sets of the whole regions only, the half-max isophotes only, and the union of them, respectively.
The shadow ribbon around each fitting line illustrates the fitting errors of $\pindex$ and $\pintercept$.
We label the fitting results around the fitting lines with corresponding text colors, below them is the value of the Pearson Correlation Coefficient (PCC).

For the whole regions, half-max isophotes, and the union of them, the fitting results of $\pindex$ are $0.40\pm 0.02, 0.38\pm 0.03$ and $0.28\pm 0.01$, while the $\pintercept$ values are $0.83\pm 0.03, 1.37\pm 0.03$, and $1.09\pm 0.02 \kps$, respectively.
Based on the PCC values of the three sample sets, 0.72, 0.49, and 0.72, we could conclude that the correlation is more obvious when the whole region samples are included due to their larger $\cloudRsky$ range. 
Figure~\ref{fig:hist-structs}(b) has shown that the half-max isophotes have very similar $\cloudveldisp$ distribution to the whole regions.
Therefore, combining the half-max isophotes with the whole regions would yield a shallower power-law relation with $\pindex=0.28\pm 0.01$.
Meanwhile, using only the half-max isophotes would have a $\pindex$ value close to that with the whole regions only.
The power-law scaling between $\cloudveldisp$ and $\cloudRsky$ is more obvious when the surface density is in a limited range, similar to the samples in \citet{1981MNRAS.194..809L} where the molecular clouds have nearly constant column densities. 
This feature is often considered an observational selection effect because the column density distribution in log-normal or power-law forms would have a mean column density close to the threshold \citep{2012MNRAS.427.2562B}.
Choosing a higher threshold to define dense subregions like the half-max isophote samples here does increase the $\cloudsurfacedensity$ range and change the relation. 
However, as discussed in Sect.~\ref{sect:spatial_scale}, the true physical scales corresponding to the measured velocity dispersion are suspicious. 
Besides, the whole region and half-max isophote samples seem like different objects. 
As we have shown in Sect.~\ref{sect:structure-properties}, the half-max isophotes match the common definition of clumps, while the whole regions are closer to (giant) molecular clouds. 
Mixing these samples would obscure the physical interpretation. 
We have already shown that the whole regions and half-max isophotes could independently have scaling relations similar to the classical result of \citet{1981MNRAS.194..809L}.
In Sect.~\ref{sect:pixelwiseresults}, we will try to conduct an analysis directly using the pixels in our sample to reduce the impact of the structure boundary and observational selection effect.

\subsection{The $\cloudveldisp$-$\cloudRsky \cdot \cloudsurfacedensity$ Relation} \label{sect:struct-surfacedensity}

\begin{figure}[t]
   \centering
   \includegraphics[width=\textwidth]{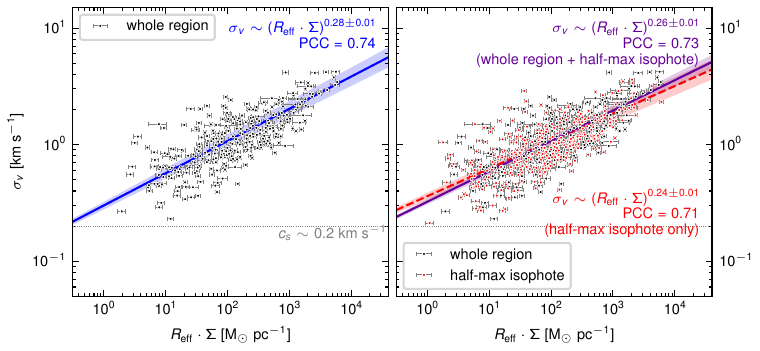}
   \caption{
   Relation between the velocity dispersion ($\cloudveldisp$) and the product of effective radius and surface density $(\cloudRsky \cdot \cloudsurfacedensity)$.
   The error bars only include the uncertainty of distance measurements. 
   Other elements are the same as Fig~.\ref{fig:larson1}.
   }
   \label{fig:vel_disp-RxSD}
\end{figure}

Compared with $\cloudveldisp$-$\cloudRsky$, the power-law relation between $\cloudRsky \cdot \cloudsurfacedensity$ and $\cloudveldisp$ seems to be a more robust one that can tolerate the mixture of samples with various surface density and galactic environments \citep[see][for a review]{2015ARA&A..53..583H}.
We explore the $\cloudveldisp$-$\cloudRsky\cdot \cloudsurfacedensity$ relation using our sample. 
Figure~\ref{fig:vel_disp-RxSD} has the same element setting as Fig.~\ref{fig:larson1} in Sect.~\ref{sect:larson1}. 
Using ODR again, we fit the power-law relation of
\begin{equation} \label{eq:heyersdR}
    \frac{\cloudveldisp}{\sigma_v'} = \left(\frac{\cloudRsky\cdot \cloudsurfacedensity}{1~\mathrm{M_{\odot}~pc^{-1}}}\right)^{\pindex'},
\end{equation}
where $\pindex'$ is the power-law index and $\sigma_v'$ is the scaling coefficient.

There is obviously less difference between using the whole regions, half-max isophotes, and the union of them than that of the $\cloudveldisp$-$\cloudRsky$ relation.
The fitting values of $\pindex'$ for the three sample sets are $0.28\pm 0.01, 0.24\pm0.01, $ and $0.26\pm 0.01$, while the values of $\sigma_v'$ are $0.30 \pm 0.02, 0.36\pm 0.02,$ and $0.32\pm 0.02 \kps$, respectively.
The $\pindex'$ value derived from the sample union is in between those of the whole regions and half-max isophotes.
Including $\cloudsurfacedensity$ in the explanatory variable naturally eliminates the impact of a large $\cloudsurfacedensity$ dynamic range that disturbs the $\cloudveldisp$-$\cloudRsky$ relation. 

Even though the $\pindex'$ derived here is smaller than the expected value of 0.5 \citep{2009ApJ...699.1092H, 2015ARA&A..53..583H}, the fact that $\cloudRsky \cdot \cloudsurfacedensity$ has a better correlation with $\cloudveldisp$ than $\cloudRsky$ is solid.
In recent studies, including galaxy-wide research \citep[e.g.][]{2017ApJ...834...57M} using the \CO{12}{}{(J=1-0)} survey of \citet{2001ApJ...547..792D} and dense gas structures \citep{2023arXiv231201497Z} with $\mathrm{H^{13}CO^{+}}{~J=1-0}$ line, also have power-law indices shallower than 0.5. 

\newcommand{\pixelwiseresults}{Pixel-wise Results}
\section{ \pixelwiseresults} \label{sect:pixelwiseresults}

This section uses the 289,696 pixels where the \CO{13}{}{} spectra have peak intensity greater than $5\sigma_{\mathrm{rms}}$ as the sample to explore the relations between velocity dispersion and other physical properties. 

\subsection{Physical Property Distributions}

\begin{figure}[t]
   \centering
   \includegraphics[width=\textwidth]{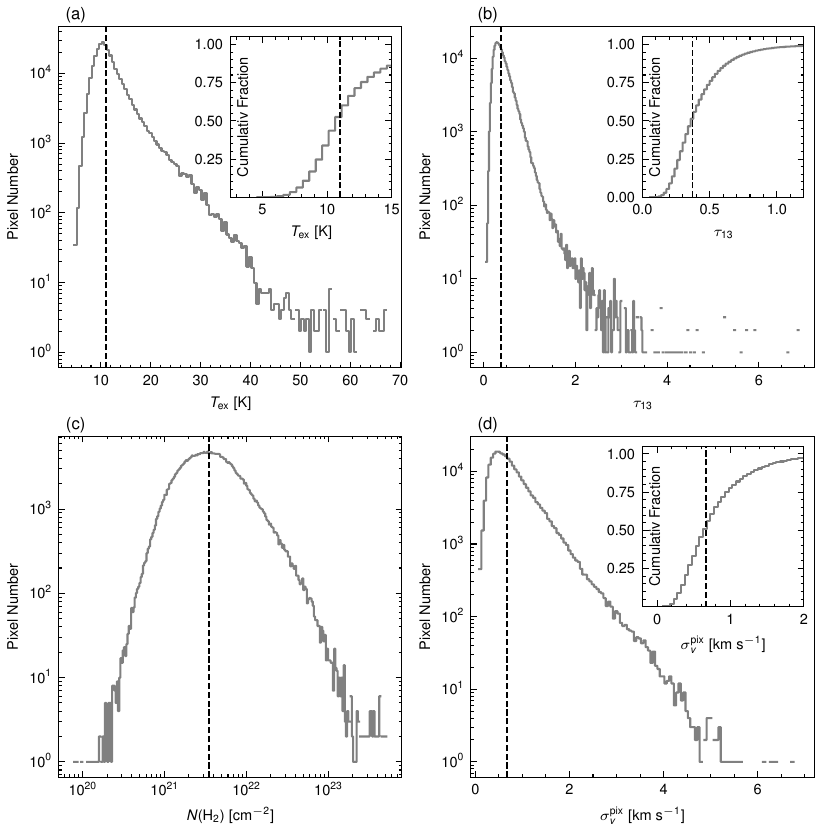}
   \caption{
      Physical property distributions of the pixels in the identified \CO{13}{}{} structures. 
      Three inset panels are created to show the cumulative probability function of $\T{ex}{}, \tau_{13},$ and $\losveldisp$.
      The vertical dashed lines in all panels denote the median values. 
   }
   \label{fig:hist-pixels}
\end{figure}

The distributions of excitation temperature ($\Tex$), central optical depth ($\tau_{13}$), $\HyMol$ column density ($N(\HyMol)$) and velocity dispersion ($\losveldisp$) measured on the pixels are demonstrated in Fig.~\ref{fig:hist-pixels}.
As shown in Fig.~\ref{fig:hist-pixels}(a), the excitation temperature derived from \T{peak}{\CO{12}{}{}} has a median value of $\sim 11~\mathrm{K}$, while the 95\% quantile is $\sim 20~\mathrm{K}$.
This distribution is consistent with the typical temperature range from $10$ to $30~\mathrm{K}$ in molecular clouds, implying thermal excitation conditions in most locations of our sample. 

Figure~\ref{fig:hist-pixels}(b) illustrates the central optical depth ($\tau_{13}$) distribution of the \CO{13}{}{} emission.
The $5\%, 50\%, $ and $95\%$ quantiles are $0.20, 0.37$, and $0.82$, respectively. 
Therefore, we could conclude that the \CO{13}{}{} emissions are optically thin in most locations of our sample and safely ignore the opacity-broadening effect. 
According to \citet{2016A&A...591A.104H}, when $\tau_{13}=0.3$ and $1.0$, the relative contribution to the total observed line width from opacity broadening are about $5\%$ and $15\%$, respectively. 

The $N(\HyMol)$ has a log-normal-like distribution shown in Fig.~\ref{fig:hist-pixels}(c), corresponding to the regime in MCs dominated by turbulence. 
Such analysis is well-known as N-PDFs \citep[e.g.,][]{2009A&A...508L..35K,2022ApJS..262...16M}.
The power-law tail associated with self-gravity is not visible here, because the tail only occupies a small fraction of the locations in each structure, and the accumulation of hundreds of structures can obscure such a feature. 
The $N(\HyMol)$ values span about three orders of magnitude. 
The median value is $3.5\times 10 ^{21}~\mathrm{cm^{-2}}$, and about $5\%$ pixels have $N(\HyMol) \lesssim 10^{21}~\mathrm{cm^{-2}}$.

Figure~\ref{fig:hist-pixels}(d) demonstrates the $\losveldisp$ distribution
with a median value of $0.667\kps$. 
Only about three thousand pixels (1\% of total) have $\losveldisp < c_s \sim 0.2\kps$, indicating that even on the small scale of pixels, the measured $\losveldisp$ are dominated by supersonic motions.
We also checked the subsonic ratio in the dropped pixels where $\T{peak}{\CO{13}{}{}} < 5\sigma_{\mathrm{rms}}$, resulting in about $7\%$. 

\subsection{Velocity Dispersion and Beam Physical Size}

The distance $(D)$ of our sample covers a large range between $400~\pc$ and $15\kpc$, making the spatial scale of the telescope beam ($\pixRsky$) varies from $\sim 0.05\pc$ to $\sim 2 \pc $.
Although the velocity measurements are limited on the line of sight, larger $\pixRsky$ could still correspond to the larger vortex of turbulent motions. 
Therefore, we use the varying $D$ and $\pixRsky$ to explore the relation with $\losveldisp$.

\begin{figure}[t!]
   \centering
   \includegraphics[width=\textwidth]{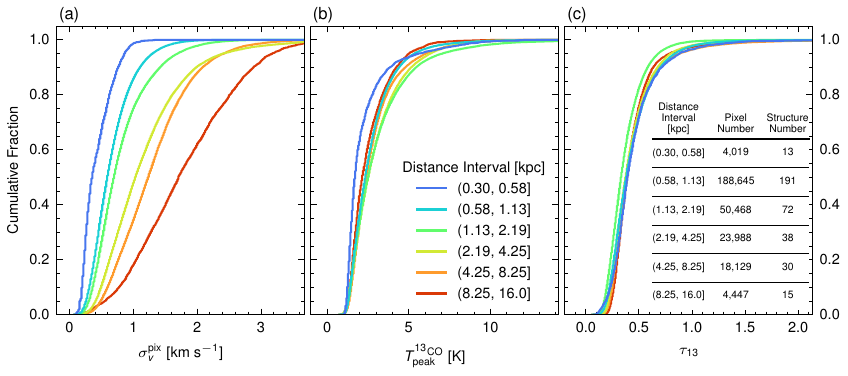}
   \caption{
      Cumulative distributions of $\losveldisp$, $\T{peak}{\CO{13}{}{}}$, and $\tau_{13}$ in logarithmically spaced $D$ intervals. 
   }
   \label{fig:pix-veldisp-D-CDF}
\end{figure}

Before explicitly correlating $\losveldisp$ with $D$ and $\pixRsky$, we investigate the distribution of $\losveldisp$ in various distance intervals. 
As shown in Fig.~\ref{fig:pix-veldisp-D-CDF}(a), the $\losveldisp$ values are split into six distance intervals.  
These intervals are logarithmically spaced.
The $D$ boundaries, pixel numbers, and structure numbers in each interval are placed in the legend and inset table of Fig.~\ref{fig:pix-veldisp-D-CDF}(b) and (c).
The cumulative distributions of $\losveldisp$ move rightward as $D$ increases. 
For comparison, the similar cumulative distributions of $\T{peak}{\CO{13}{}{}}$ and $\tau_{13}$ are shown in Fig.~\ref{fig:pix-veldisp-D-CDF}(b) and (c), respectively.
Unlike $\losveldisp$, the distributions of $\T{peak}{\CO{13}{}{}}$ and $\tau_{13}$ show less dependence on $D$. 

\begin{figure}[t!]
   \centering
   \includegraphics[width=\textwidth]{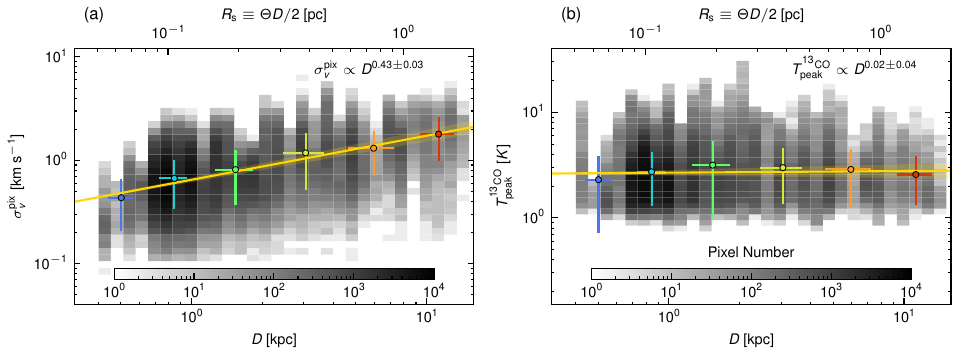}
   \caption{
      Histograms and power-law fitting results of $\losveldisp$-$D$ and $\T{peak}{\CO{13}{}{}}$-$D$.
      The data points and error bars are the mean and standard derivation values within the same distance intervals as Fig.~\ref{fig:pix-veldisp-D-CDF}.
    }
   \label{fig:pix-veldisp-D-fit}
\end{figure}

To explicitly show the relation between $\losveldisp$ and $D$, we plot the histograms as the background of Fig.~\ref{fig:pix-veldisp-D-fit}(a).
Even though the $\losveldisp$ values are vertically scattered, the trend of $\losveldisp$ increasing with $D$ is visible. 
The six data points with error bars use the same $D$ interval configuration as Fig.~\ref{fig:pix-veldisp-D-CDF}.
Each data point is derived from the mean $\losveldisp$ and $D$ values of pixels in each $D$ interval.
Correspondingly, the error bars are the standard derivations. 
Using these points, we fit a power-law relation with ODR. 
The gold line in Fig.~\ref{fig:pix-veldisp-D-fit}(a) is the fitting result of $\losveldisp \propto D^{0.43 \pm 0.03}$.

We also apply the same operations on $\T{peak}{\CO{13}{}{}}$ and show the results in Fig.~\ref{fig:pix-veldisp-D-fit}(b).
The bottom part of the distribution corresponds to the $5\sigma_{\mathrm{rms}}$ threshold described at the end of Sect.~\ref{sect:property}, while the top part fluctuates with different structures. 
In contrary to $\losveldisp$, $\T{peak}{\CO{13}{}{}}$ shows no obvious trend with $D$. 
The ODR fitting result is $\T{peak}{\CO{13}{}{}}\propto D^{0.02\pm 0.04}$, indicating no dependence on $D$.
Therefore, we could attribute the increasing $\losveldisp$ to the increasing $D$. 
\citet{2020A&A...633A..14R,2020A&A...640A..72R} referred to this phenomenon as the beam averaging effect.
Using the sample with accurate distance measurements, we have acquired the power-law relation of $\losveldisp \propto D^{0.43 \pm 0.03}$, or equivalently, $\losveldisp \propto \pixRsky^{0.43 \pm 0.03}$.

Even though these two quantities are measured in different spatial directions, i.e., $\pixRsky$ on the plane-of-sky and $\losveldisp$ along the line-of-sight, their power-law correlation might still reflect the turbulent nature of molecular gas in the ISM. 
A turbulent flow contains eddies across different spatial scales, where larger spatial separation corresponds to eddies with larger scale and higher velocity difference \citep[See figure 1 in][for a schematic illustration]{2024arXiv240810406V}.
For eddies smaller than the beam's physical size, the corresponding velocity difference is unresolved and contributes to the line width/velocity dispersion measured on the pixel. 
The trend shown in Fig.~\ref{fig:pix-veldisp-D-fit}(a) might statistically reflect such features of turbulent molecular gas. 
Nevertheless, the large vertical scatter indicates other important factors determining the measured $\losveldisp$ on each pixel, e.g., the column density (Sect.~\ref{sect:pix-veldisp-NH2}). 

\begin{figure*}
    \centering
    \includegraphics[width=\textwidth]{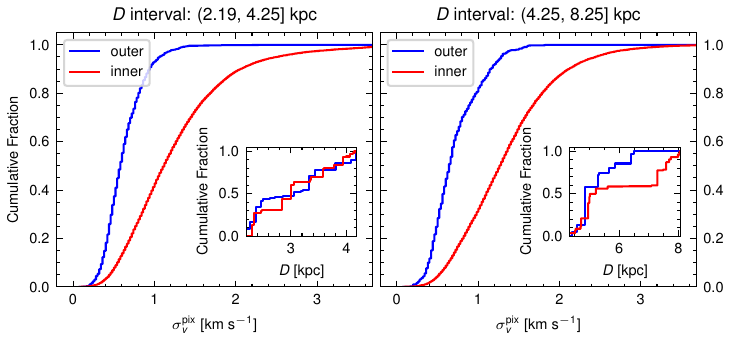}
    \caption{Cumulative distributions of $\losveldisp$ in the inner and outer galaxy. The inset panels contain the corresponding heliocentric distance ($D$) distributions. 
    We consider the samples with galactocentric radii $R_{\mathrm{gal}} < 8.15~\mathrm{kpc}$ as the inner subsets, while the rest constitute the outer ones. 
    The $\losveldisp$ mean values and standard deviations are shown in Table~\ref{table:inner_outer}.
    }
    \label{fig:pix-veldisp-inner-outer}
\end{figure*}

Besides the different beam physical sizes introduced by varying distances, the galactic environments might also play a role.  
If the clouds in the inner galaxy have stronger turbulence, the velocity dispersion should be statistically larger than the outer Galaxy.
Using the pixel-wise sample same as in Figs.~\ref{fig:pix-veldisp-D-CDF} and \ref{fig:pix-veldisp-D-fit}, the Pearson's correlation coefficient between $\losveldisp$ and the Galactocentric radii ($R_{\mathrm{gal}}$) is $-0.47$. 
Even though this value indicates a moderate correlation, we must remember that $R_{\mathrm{gal}}$ strongly depends on $D$ and the coefficient is $-0.6$.
To have a reliable assessment of the $R_{\mathrm{gal}}$ impact on $\losveldisp$, we must rule out the dependence on $D$.
Among the six distance intervals shown in Fig.~\ref{fig:pix-veldisp-D-CDF}, we noticed $(2.19, 4.25]$ and $(4.25, 8.25]~\mathrm{kpc}$ are suitable for this because the correlation coefficients between $R_{\mathrm{gal}}$ and $D$ therein are only $-0.07$ and $0.04$, respectively, much smaller than the corresponding values between $R_{\mathrm{gal}}$ and $\losveldisp$ of $-0.33$ and $-0.19$, indicating the weak negative correlations between $\losveldisp$ and $R_{\mathrm{gal}}$ in these two distance intervals. 

\begin{table}[h!]
\centering
\begin{threeparttable}
    \caption{Mean values of $\losveldisp$ in the inner and outer Galaxy}
    \label{table:inner_outer}
    \centering
    \begin{tabularx}{0.7\textwidth}{
    @{}
    >{\centering\arraybackslash}X
    >{\centering\arraybackslash}X
    >{\centering\arraybackslash}X
    @{}
    }
    \hline
Distance interval (kpc) & Inner         & Outer      \\ \hline
(2.19, 4.25{]}    & $1.26(0.67)$  &  $0.63(0.24)$ \\
(4.25, 8.25{]}    &$1.35(0.60)$  & $0.72(0.30)$ \\ \hline
\end{tabularx}
\begin{tablenotes}
\footnotesize
\item (a) Values in the parentheses are the corresponding standard deviations;
\item (b) The boundary between the inner and outer Galaxy is set as $R_\mathrm{gal}=8.15~\mathrm{kpc}$;
\item (c) The distance intervals are heliocentric while $R_{\mathrm{gal}}$ is the galactocentric radius. 
\end{tablenotes}
\end{threeparttable}
\end{table}

To visually show the $\losveldisp$ distribution difference between the inner and outer Galaxy, we plot the cumulative distributions in Fig.~\ref{fig:pix-veldisp-inner-outer}. 
The pixel-wise $D$ distributions are also demonstrated in the inset panels to evaluate the $D$ variation.
In both distance intervals, $\losveldisp$ distributions in the inner and outer Galaxy are significantly different. 
The $\losveldisp$ values are statistically larger in the inner part. 
Beside Fig.~\ref{fig:pix-veldisp-inner-outer}, this result is also demonstrated through the mean $\losveldisp$ values in Table~\ref{table:inner_outer}.
Because the adopted distances in our sample are based on the parallaxes of masers or background stars (Sect.~\ref{sect:distance}), the above analysis does not encounter the problem of kinematic distance ambiguity at the cost of a smaller sample size. 
The two selected distance intervals contain only 38 and 30 gas structures, respectively, making the heliocentric distance CDFs in a stepped shape. 
Even though this does not affect the result that the pixel-wise velocity dispersions are systematically larger in the inner galaxy, a more robust conclusion should be made from a larger sample with more completeness, which is mostly limited by the distance measurements for now.

\subsection{Velocity dispersion and Column Density} \label{sect:pix-veldisp-NH2}

Previous works have concluded the importance of the surface density ($\cloudsurfacedensity$).
When $\cloudsurfacedensity$ is included, changing the spatial boundaries of the structures does not significantly affect the scaling relation.
To increase the dynamic range of $\cloudsurfacedensity$, we followed \citet{2009ApJ...699.1092H} to define the subregions within the $N(\HyMol)$ half-max isophote in the previous section.
As the velocity dispersion ($\losveldisp$) and column density $N(\HyMol)$ on each pixel of our sample have been measured, we could directly correlate the $\losveldisp$ and $N(\HyMol)$ values.
In this manner, the column density range traceable by \CO{13}{}{} emission can be fully utilized. 
Here we first demonstrate the relation between $\losveldisp$ and $N(\HyMol)$ using the mixture of pixels from all structures, then statistically analyze the relation in each structure. 

\begin{figure*}[h!]
   \centering
   \includegraphics[width=0.75\textwidth]{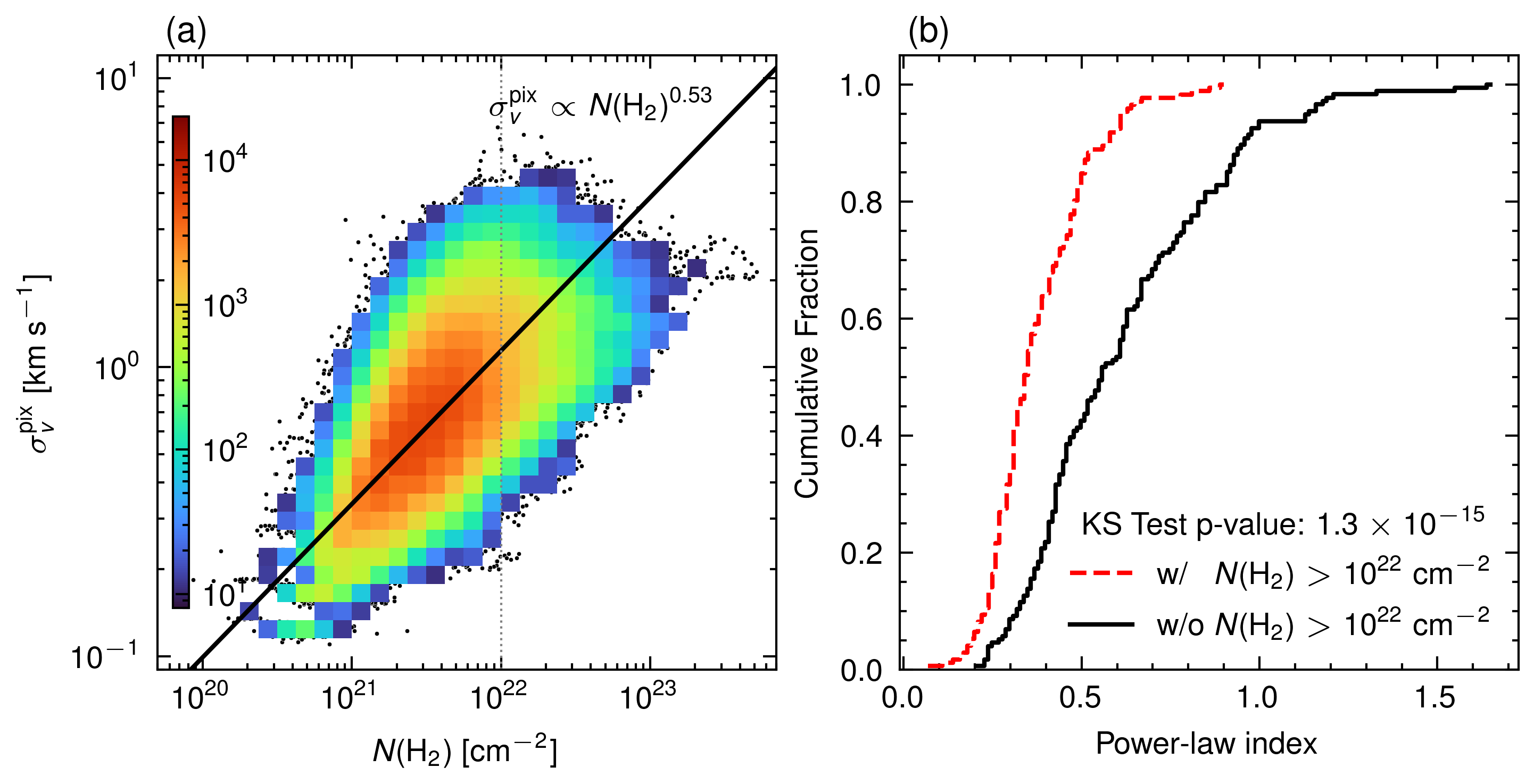}
   \caption{
      Panel (a): Two-dimensional histogram of $\losveldisp$ against $N(\HyMol)$. 
      The color bar represents the number of samples. 
      The scatter points replace the 2D bins with less than ten samples.  
      Panel (b): The power-law index distributions of  $\losveldisp$-$N(\HyMol)$ relation in each structure. 
      The structures are divided into two subsets according to the existence of pixels with $N(\HyMol) > 10^{22}~\mathrm{cm^{-2}}$ therein. 
      The p-value of the Kolmogorov-Smirnov test indicates the two subsets have significantly different power-law indices. 
   }
   \label{fig:pixel-NH2}
\end{figure*}

Figure~\ref{fig:pixel-NH2}(a) displays the correlation between $\losveldisp$ and $N(\HyMol)$ of pixels in all structures using 2D-histograms as the background, where $\losveldisp$ is positively correlated with $N(\HyMol)$ and the Pearson correlation coefficient is 0.65.
With such a strong correlation in the log-log plot, it is straightforward to fit a power-law relation in the form of $\losveldisp \propto N(\HyMol)^{\xi}$.
The ODR fitting result is $\losveldisp \propto N(\HyMol)^{0.53}$. 
Even though the power-law index ($\xi$) is close to a common value of 0.5, we must note that the $\losveldisp$ and $N(\HyMol)$ values were measured on each line-of-sight, instead of the complete cloud structures.
Therefore, interpreting this relation should involve the gas structure thickness, a hidden variable in most observations. 
As the column density is the integral of volume density along the line-of-sight, $N(\HyMol)$ comprises the thickness and volume density profile. 
On the other hand, $\losveldisp$ is a direct measurement of the particle velocity distribution in the same direction. 
The relation shown in Fig.~\ref{fig:pixel-NH2}(a) might be interpreted by the turbulent motion scaling with the line-of-sight thickness.

\begin{figure*}[h!]
   \centering
   \includegraphics[width=\textwidth]{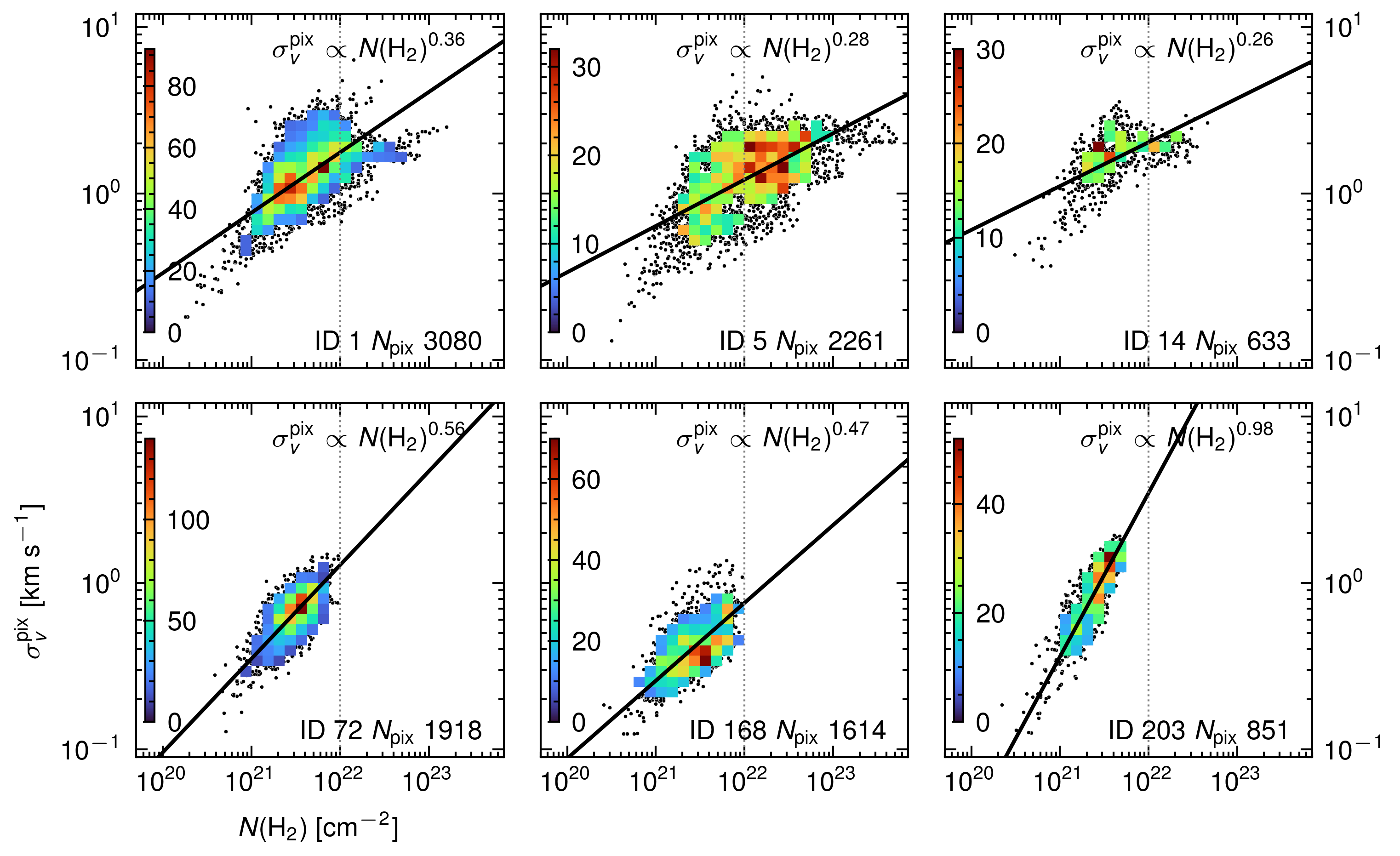}
   \caption{
      Examples of the 2D histograms between $\losveldisp$ and $N(\HyMol)$. 
      Each panel contains one structure, where the structure ID and pixel number above $5\sigma_{\mathrm{rms}}$ therein are labeled at the bottom. 
      The three structures on the top row contain pixels with $N(\HyMol) > 10^{22}~\mathrm{cm^{-2}}$, while the other three at the bottom row do not. 
      The other elements in each panel are the same as Fig.~\ref{fig:pixel-NH2}(a).
   }
   \label{fig:pixel-NH2-by-struct}
\end{figure*}

The largest $\losveldisp$ values locate around $N(\HyMol)\sim 10^{22}~\mathrm{cm^{-2}}$, above which the increasing of $\losveldisp$ slows down and seems to be flat around $\losveldisp \sim 2~\mathrm{km~s^{-1}}$. 
To further explore such a pattern, we replicate Fig.~\ref{fig:pixel-NH2}(a) for each structure.  
Some examples are shown in Fig.~\ref{fig:pixel-NH2-by-struct}, representing two kinds of $\losveldisp$-$N(\HyMol)$ distributions in our sample. 
One kind is shown in the top row of Fig.~\ref{fig:pixel-NH2-by-struct}, similar to the trend in Fig.~\ref{fig:pixel-NH2}(a), the $\losveldisp$ is flatting after $N(\HyMol)\gtrsim 10^{22}~\mathrm{cm^{-2}}$.
The other kind can be well described by power-law relations, as shown by the bottom row of Fig.~\ref{fig:pixel-NH2}.
As the flat trend at the high end of column density would decrease the power-law index, we expect that there would be a difference on $\xi$ distribution between the structures with and without high-column-density pixels. 
Figure~\ref{fig:pixel-NH2}(b) shows the cumulative distributions of the power-law indices ($\xi$) between $\losveldisp$ and $N(\HyMol)$ in the two structure groups.
The mean (std) value of $\xi$ in the structures with high-column-density pixels is $0.38~(0.14)$, while in the group without such pixels, it is $0.62~(0.27)$.
The Kolmogorov-Smirnov test between the two groups'  $\xi$ distributions has a p-value of $1.3\times 10^{-15}$, rejecting the null hypothesis of identical distributions.  
The numbers of structures with and without high-column-density pixels  are 174 and 171, respectively, where the structures with less than $16$ pixels above 5$\sigma_{\mathrm{rms}}$ have been dropped to have reliable fitting results.

The interesting $\losveldisp$ flattening pattern at the high end of $N(\HyMol)$ is similar to the velocity dispersion profile analysis conducted by \citet{2023MNRAS.525.2935P}. 
They found the radial profiles of $\losveldisp$ are different in the diffuse and denser parts of most clouds. 
The flat $\losveldisp$ profile in the denser part is consistent with the trend in our structures with high-column-density pixels , e.g., the examples in the top row of Fig.~\ref{fig:pixel-NH2-by-struct}. 
A possible interpretation of these features is   that the embedded dense structures (clumps or cores) have been decoupled from the surrounding gas structure and dominated by gravitational contraction, in the meanwhile, the parent clouds could still be stable and supported by turbulence.
Such structures in our sample have a viral parameter median value $\sim 2$ (Sect.~\ref{sect:virial}), also supporting this scenario. 
It could be consistent with the quasi-isolated gravitational collapse model proposed by \citet{2017MNRAS.465..667L}.
However, we must note that this interpretation of our observational results is still primitive. 
More evidence and analytical proof are required to establish a solid connection. 

The $\losveldisp$-$N(\HyMol)$ relations of the  structures with only low-column-density pixels can be well described by the power-law form, as shown in the bottom row of Fig.~\ref{fig:pixel-NH2-by-struct}.
The mean standard deviation of the residuals is $\lesssim 0.1~\mathrm{dex}$.
As we will discuss in Sect.~\ref{sect:spatial_scale}, this strong power-law correlation might be interpreted as the turbulent motions scaling with the line-of-sight thickness of the molecular gas. 
Beside turbulence, gravity might also play a role in these relations because the increasing $N(\HyMol)$ includes more amount of gas.
As it is still under debate whether the measured velocity dispersion represents the turbulent flow that supports the gas structure against gravitational contraction, the systematic infall motions during the collapsing, or even the mixture of both \citep{2024arXiv240810406V}, 
the above interpretation of the power-law relation between $\losveldisp$ and $N(\HyMol)$ is not conclusive for now. 

\section{Discussion} 
\label{sect:discuss}

\subsection{The Keto-Heyer Diagram and Virial Parameter}
\label{sect:virial}

\begin{figure}
   \centering
   \includegraphics[width=1.0\textwidth]{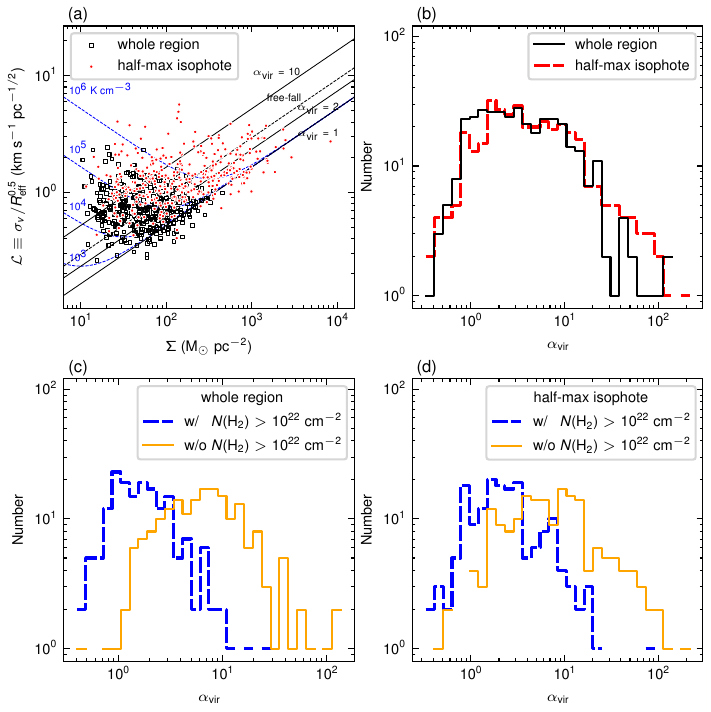}
   \caption{
   Panel (a): Dependence of the Larson Ratio $\mathcal{L} \equiv \cloudveldisp / \cloudRsky^{0.5}$ on the surface density $\Sigma$.
      The black squares and red circles denote the whole regions and $N(\HyMol)$ half-max isophotes of the identified \CO{13}{}{} structures. 
      The solid straight lines show the loci of $\virparam=1, 2, 10$ while the dashed one corresponds to the free-fall interpretation  \citep{2011MNRAS.411...65B}. 
      The blue dashed lines in V-shapes indicate the loci of pressure-bounded virial equilibrium  \citep[PVE,][]{2011MNRAS.416..710F}  with various external pressure $P_{\mathrm{e}}/k_{\mathrm{B}}$. 
      Panel (b): Virial Parameter ($\virparam$) distributions of the whole regions and half-max isophotes. 
      Panels (c) and (d): the $\virparam$ differences between structures with and without high-column-density pixels where $N(\HyMol) > 10^{22}~\mathrm{cm^{-2}}$.
}
   \label{fig:larsonratio}
\end{figure}

Here we discuss the dynamic state of the \CO{13}{}{} structures through the Keto-Heyer diagram \citep[KH diagram,][]{1986ApJ...304..466K,2009ApJ...699.1092H} and virial parameters.  The KH diagram correlates the Larson ratio $\mathcal{L}\equiv \cloudveldisp / \cloudRsky^{0.5}$ with the surface density ($\cloudsurfacedensity$).
The viral parameter ($\virparam$), surface density ($\cloudsurfacedensity$), and $\mathcal{L}$ can be related with 
\begin{equation}
    \mathcal{L} = \left(\frac{\pi \virparam G \cloudsurfacedensity}{5}\right)^{\frac{1}{2}},
\end{equation}
in which the viral parameter ($\virparam$) defined by \citet{1992ApJ...395..140B} is
\begin{equation} \label{eq:vir}
    \virparam \equiv \frac{2E_k}{|E_g|} \approx \frac{5\cloudveldisp ^2 \cloudRsky }{G\cloudMass} = \frac{5\cloudveldisp^2}{\pi G \cloudRsky \cloudsurfacedensity}.
\end{equation}
Using the physical properties derived in the whole regions and half-max isophotes of our sample (Sect.~\ref{sect:property}), we plot the KH diagram in Fig.~\ref{fig:larsonratio}(a). 
We also include the loci of $\virparam = 1, 2, 10$ and pressure-bound virial equilibrium with external pressure $P_{\mathrm{e}} / k_\mathrm{B} \sim 10^3$ to $10^{6} \Kelvin~\mathrm{cm^{-3}}$ in Fig.~\ref{fig:larsonratio}(a).
The pressure-bounded virial equilibrium \citep[PVE,][]{2011MNRAS.416..710F} can be described with
\begin{equation}
    \mathcal{L}^{2} = \frac{1}{3}\left(\pi \Gamma \cloudsurfacedensity + \frac{4P_\mathrm{e}}{\cloudsurfacedensity}\right),
\end{equation}
where $P_{\mathrm{e}}$ is the external pressure and $\Gamma=3/5$ is a form factor for a sphere of constant density. 

It seems the sample distribution in Fig.~\ref{fig:larsonratio}(a) is consistent with the prediction that $\mathcal{L}\propto \cloudsurfacedensity^{\frac{1}{2}}$. 
However, this trend is only obvious when the half-max isophote samples are included.
Removing them would make the correlation vanish. 
Specifically, the Pearson correlation coefficients between $\mathcal{L}$ and $\cloudsurfacedensity$ with and without half-max isophote samples are $0.39$ and $-0.02$, respectively. 
As we have addressed in Sect.~\ref{sect:structure-properties}, the $\cloudsurfacedensity$ values of the whole regions are limited in the range of $[10.6, 460]~\Msunpersqpc$, while the half-max isophotes can increase the upper bound of $\cloudsurfacedensity$ to $8.4\times 10^{3}~\Msunpersqpc$.
Therefore, the dynamic range of $\cloudsurfacedensity$ is critical in analyzing the $\mathcal{L}$-$\cloudsurfacedensity$ relation. 

Some super-viral structures lie above the locus of $\virparam = 10$ on the left-hand side of Fig.~\ref{fig:larsonratio}(a). 
Most of them are low-$\cloudsurfacedensity$ structures with $\cloudsurfacedensity < 100~\Msunpersqpc$. 
The interpretations of such super-virial objects are under debate. 
On the one hand, if the molecular clouds and clumps are supported by turbulence and in equilibrium, these super-virial structures are over-supported and require external pressure to be confined \citep[e.g.,][]{1992ApJ...395..140B, 2011MNRAS.416..710F}. 
The V-shaped curves in Fig.~\ref{fig:larsonratio}(a) also illustrate that the external pressure in the order of magnitude of $\sim 10^{5}~\mathrm{K}$ might be sufficient to confine the super-viral structures. 
On the other hand, they can also be interpreted as the early evolution stage of turbulent collapsing structures \citep{2018MNRAS.479.2112B, 2019MNRAS.490.3061V}. 

The distributions of $\virparam$ in different categories could be interesting. 
To assess the impact of spatial boundary definition on $\virparam$, 
we plotted the $\virparam$ histograms of the whole regions and half-max isophotes in Fig.~\ref{fig:larsonratio}(b), and found no obvious differences between them, while the median $\virparam$ values are $3.2$ and $3.7$, respectively. 
When the spatial boundaries are shrinking from the whole region to half-max isophote, $\cloudRsky$ is heavily reduced but $\cloudsurfacedensity$ significantly increases, leading to similar $\virparam$ distributions. 
Therefore, the spatial boundary definitions have a relatively small impact.

However, as shown in Fig.~\ref{fig:larsonratio}(a), the structures with high-$\cloudsurfacedensity$ seem to have smaller $\virparam$ values.
This inspires us to use the classification based on the existence of high-column-density ($> 10^{22}~\mathrm{cm^{-2}}$) pixels in Sect.~\ref{sect:pix-veldisp-NH2}.
As shown in Fig.~\ref{fig:larsonratio}(c), the $\virparam$ distributions of the whole regions are different between the structures with and without such high-column-density pixels, where the median $\virparam$ values are 1.7 and 6.7, respectively. 
As demonstrated in Fig.~\ref{fig:larsonratio}(d), the half-max isophotes also have the same trend, while the corresponding median $\virparam$ values are 2.2 and 6.7. 
In Sect.~\ref{sect:pix-veldisp-NH2}, we have found that the structures with and without high-column-density pixels have different $\losveldisp \propto N(\HyMol)^{\xi}$ relations.
Now we know they also have different $\virparam$ distributions.

\subsection{Spatial Scale of the Measured Velocity Dispersion} 
\label{sect:spatial_scale}
A prominent feature of interstellar turbulence in the inertia range is the power-law correlation between the spatial scale and velocity difference.
In observations, the measurement of velocity along the line-of-sight is much easier than the estimation of spatial scales, which is why we have only analyzed the structures with distance measurements (Sect.~\ref{sect:distance}) at the cost of a smaller sample size. 
An empirical assumption in the velocity dispersion-size relation is that the structure thickness along the line-of-sight (LoS) is comparable to its plane-of-the-sky (PoS) size. 
However, definitions of spatial boundaries are rather arbitrary and depend on the choice of emission line, data sensitivity, intensity thresholds, and structure identification methods. 
Meanwhile, the LoS velocity measurements are less affected, e.g., the comparison of $\cloudveldisp$ distributions between the whole region and half-max isophote samples in Sect.~\ref{sect:structure-properties}, shrinking the spatial boundaries toward the structure density center does not significantly change the $\cloudveldisp$ distribution.
If we accept the comparable LoS-PoS size assumption, the thickness values of the half-max isophotes should be closer to the PoS sizes of their parental whole regions. 
Therefore, including the half-max isophotes in Fig.~\ref{fig:larson1}(b) might naturally break the power-law relation. 

There is other evidence implying the importance of the hidden LoS thickness.
For instance, the $\cloudveldisp$-$\cloudRsky \cdot \cloudsurfacedensity$ relations shown in Fig.~\ref{fig:vel_disp-RxSD} are more robust than the $\cloudveldisp$-$\cloudRsky$ relations in Fig.~\ref{fig:larson1}. 
The participation of the half-max isophote samples does not significantly impact the $\cloudveldisp$-$\cloudRsky \cdot \cloudsurfacedensity$ relation. 
Because the surface/column density is the LoS integral of the volume density, it could be a proxy of the LoS thickness. 
Comparing with the PoS-only $\cloudRsky$, $\cloudRsky \cdot \cloudsurfacedensity$ includes the extra information of the LoS thickness and might be a rough 3D estimation of the spatial scale, which corresponds better with $\cloudveldisp$ than the PoS-only $\cloudRsky$.
Another piece of evidence is the pixel-wise $\losveldisp$-$N(\HyMol)$ relations in Sect.~\ref{sect:pix-veldisp-NH2}.
As shown in Fig.~\ref{fig:pixel-NH2}(a) and examples in Fig.~\ref{fig:pixel-NH2-by-struct}, the power-law relations between $\losveldisp$ and $N(\HyMol)$ below $10^{22}~\mathrm{cm^{-2}}$ are quite obvious. 
The indices around $0.5$, either with the pixel mixture of all structures or the statistical results from the fitting in each structure, could be the consequence of the interstellar turbulence. 
Besides, if we assume the average volumetric number density along the LoS in a  structure without high-column-density pixels is close to the \CO{13}{}{} critical density, the column density $N(\HyMol)$ is directly proportional to the LoS size.
Then the mean power-law index of $0.62$  is consistent with the value of $3/5$ proposed by \citet{2021ApJ...906L...4C}.

A more reasonable spatial scale estimation of the measured velocity dispersion should be some form of average between the LoS thickness and PoS size. 
However, as a dimension hidden by projection effects in astronomical observations, it is inherently hard to measure the thickness.
Sometimes the thickness could also be surprising. 
For example, an edge-on sheet looks like a filament from our angle, such as the Musca molecular cloud \citep{2018Sci...360..635T,2022MNRAS.514.3593T}.
In previous studies, some methods have been developed to estimate the thickness of molecular clouds along the LoS,
for example, 3D differential extinction maps \citep[e.g.,][]{2021ApJ...919...35Z,2022MNRAS.514.3593T}, radiation transfer \citep[e.g.,][]{2012ApJ...756...12L}, and core velocity dispersion \citep[CVD,][]{2015ApJ...811...71Q}.
The 3D extinction maps are restricted to the solar neighborhood, while the CVD relies on models. 
The most promising approach is estimating the volume density using radiation transfer and emission lines with multi-transition observations. However, collecting such a complex data set with enough spatial coverage is not trivial. 

\section{Summary} \label{sect:summary}

The physical interpretations of the relation between velocity dispersion and spatial size of molecular cloud structures are still under debate, 
even though it has been more than four decades since the first realization \citep{1981MNRAS.194..809L}.
People keep visiting this topic for its indispensable function in revealing the dynamic of molecular cloud structures.
In this work, we applied the ISMGCC method on the MWISP CO data and acquired 360 \CO{13}{}{} structures with accurate distance measurements to analyze the correlations between the physical properties therein.  
Here we summarize our results: 
\begin{enumerate}
   \item The velocity dispersion and effective radius of the entire structures show a scaling relation of $\cloudveldisp \propto \cloudRsky^{0.40\pm 0.02}$, while including the $N(\HyMol)$ half-max regions would result in a shallower scaling of $\cloudveldisp \propto \cloudRsky^{0.28\pm 0.01}$.
   This indicates that the $\cloudveldisp$-$\cloudRsky$ relation is less solid when the sample spans a larger $\cloudsurfacedensity$ range, consistent with the conclusion from  previous works;  
   \item The $\cloudveldisp$-$\cloudRsky \cdot \cloudsurfacedensity$ is less sensitive to the $\cloudsurfacedensity$ range and has a power-law index below 0.3;
   \item We also analyzed the pixels above $5\sigma_{\mathrm{rms}}$ in the structures and noticed a scaling relation of $\losveldisp \propto D^{0.43\pm 0.03}$ or $\pixRsky^{0.43\pm 0.03}$. The possible interpretation is that the larger beam physical sizes ($\cloudRsky$) with increasing distances would cover larger turbulent eddies, statistically resulting in higher velocity dispersion. Besides, $\losveldisp$ in the inner galaxy seems statistically larger than the outer side.
   \item The pixel-wise $\losveldisp$ has stronger correlation with $N(\HyMol)$ than with $\pixRsky$, in a power-law form of $\losveldisp \propto N(\HyMol)^{\xi}$  .
   Fitting the relation with the mixture of pixels from all structures results in $\xi = 0.53$.
   Each structure could have its own $\xi$ value through fitting the relation with its pixels.  
   We noticed a $\xi$ distribution difference between the structures with and without high-column-density ($>10^{22}~\mathrm{cm^{-22}}$) pixels, where the mean (std) $\xi$ values are $0.38 (0.14)$ and $0.62(0.14)$, respectively. 
   The significant difference is mostly caused by the flattening of $\losveldisp$ when $N(\HyMol) \gtrsim 10^{22}~\mathrm{cm^{-2}}$.
   Even though the role of gravity is still unclear in the structures without high-column-density pixels, the stronger power-law correlations might reflect the turbulent nature of the diffuse molecular gas because $N(\HyMol)$ is partially a proxy of the line-of-sight thickness.
   \item The above structures with and without high-column-density pixels also have different virial parameter distributions where the former ones have median $\virparam\sim 2 $, while the latter is $\sim 7$. 
   Changing the spatial boundaries from the entire structures to the subregions within its $N(\HyMol)$ half-max isophote does not significantly affect the $\virparam$ distribution, but increases the dynamic range of the surface density ($\cloudsurfacedensity)$, making the sample distribution roughly consistent with the relation of $\cloudveldisp / \cloudRsky^{\frac{1}{2}} \propto \cloudsurfacedensity^{\frac{1}{2}}$.
\end{enumerate}

\normalem
\begin{acknowledgements}
We are deeply obliged to the MWISP working group, Xin Zhou, Yan Sun, Jixian Sun, Dengrong Lu and Binggang Ju, and the observation assistants of the project, who have been working hard in instrument maintenance, taking and reducing the data, without which this work cannot be done. 
This work is supported by the National Key R\&D Program of China (Grant No. 2023YFA1608000) and the National Natural Science Foundation of China (NSFC, Grant Nos. U2031202, 12373030, and 11873093).
MWISP is sponsored by the National Key R\&D Program of China with grant 2023YFA1608000 and the CAS Key Research Program of Frontier Sciences with grant QYZDJ-SSW-SLH047.
Z. Chen acknowledges the Natural Science Foundation of Jiangsu Province (grants No. BK20231509).
H. Feng would like to express the thanks to 
Hongchi Wang, Xuepeng Chen, Pak Shing Li, Chong Li,  Min Fang, and the anonymous reviewer for constructive suggestions that enable improvement of the manuscript.
\end{acknowledgements}
  
\bibliographystyle{raa}
\bibliography{bib}

\begin{thebibliography}{80}
\providecommand\natexlab[1]{#1}
\providecommand\JournalTitle[1]{#1}

\bibitem[{Ballesteros-Paredes} {et~al.}(2012)]{2012MNRAS.427.2562B}
{Ballesteros-Paredes}, J., {D'Alessio}, P., \& {Hartmann}, L. 2012, \mnras, 427, 2562

\bibitem[{Ballesteros-Paredes} {et~al.}(2011)]{2011MNRAS.411...65B}
{Ballesteros-Paredes}, J., {Hartmann}, L.~W., {V{\'a}zquez-Semadeni}, E., {Heitsch}, F., \& {Zamora-Avil{\'e}s}, M.~A. 2011, \mnras, 411, 65

\bibitem[{Ballesteros-Paredes} {et~al.}(2018)]{2018MNRAS.479.2112B}
{Ballesteros-Paredes}, J., {V{\'a}zquez-Semadeni}, E., {Palau}, A., \& {Klessen}, R.~S. 2018, \mnras, 479, 2112

\bibitem[{Ballesteros-Paredes} {et~al.}(2020)]{2020SSRv..216...76B}
{Ballesteros-Paredes}, J., {Andr{\'e}}, P., {Hennebelle}, P., {et~al.} 2020, \ssr, 216, 76

\bibitem[{Bensch} {et~al.}(2001)]{2001A&A...366..636B}
{Bensch}, F., {Stutzki}, J., \& {Ossenkopf}, V. 2001, \aap, 366, 636

\bibitem[{Bergin} \& {Tafalla}(2007)]{2007ARA&A..45..339B}
{Bergin}, E.~A., \& {Tafalla}, M. 2007, \araa, 45, 339

\bibitem[{Bertoldi} \& {McKee}(1992)]{1992ApJ...395..140B}
{Bertoldi}, F., \& {McKee}, C.~F. 1992, \apj, 395, 140

\bibitem[{Bian} {et~al.}(2022)]{2022AJ....163...54B}
{Bian}, S.~B., {Xu}, Y., {Li}, J.~J., {et~al.} 2022, \aj, 163, 54

\bibitem[{Bian} {et~al.}(2024)]{2024AJ....167..267B}
{Bian}, S.~B., {Wu}, Y.~W., {Xu}, Y., {et~al.} 2024, \aj, 167, 267

\bibitem[{Cai} {et~al.}(2021)]{2021RAA....21..304C}
{Cai}, J.-J., {Yang}, J., {Zheng}, S., {et~al.} 2021, Research in Astronomy and Astrophysics, 21, 304

\bibitem[{Caselli} \& {Myers}(1995)]{1995ApJ...446..665C}
{Caselli}, P., \& {Myers}, P.~C. 1995, \apj, 446, 665

\bibitem[{Cen}(2021)]{2021ApJ...906L...4C}
{Cen}, R. 2021, \apjl, 906, L4

\bibitem[{Dame} {et~al.}(2001)]{2001ApJ...547..792D}
{Dame}, T.~M., {Hartmann}, D., \& {Thaddeus}, P. 2001, \apj, 547, 792

\bibitem[{Downes} {et~al.}(2023)]{2023MNRAS.519.5427D}
{Downes}, T.~P., {Hartigan}, P., \& {Isella}, A. 2023, \mnras, 519, 5427

\bibitem[{Draine}(2011)]{2011piim.book.....D}
{Draine}, B.~T. 2011, {Physics of the Interstellar and Intergalactic Medium}

\bibitem[{Feng} {et~al.}(2024)]{2024arXiv240901181F}
{Feng}, H., {Chen}, Z., {Jiang}, Z., \& {Urquhart}, J.~S. 2024, arXiv e-prints, arXiv:2409.01181

\bibitem[{Field} {et~al.}(2011)]{2011MNRAS.416..710F}
{Field}, G.~B., {Blackman}, E.~G., \& {Keto}, E.~R. 2011, \mnras, 416, 710

\bibitem[{Frerking} {et~al.}(1982)]{1982ApJ...262..590F}
{Frerking}, M.~A., {Langer}, W.~D., \& {Wilson}, R.~W. 1982, \apj, 262, 590

\bibitem[{Fuller} \& {Myers}(1992)]{1992ApJ...384..523F}
{Fuller}, G.~A., \& {Myers}, P.~C. 1992, \apj, 384, 523

\bibitem[{Gaia Collaboration} {et~al.}(2016)]{2016A&A...595A...1G}
{Gaia Collaboration}, {Prusti}, T., {de Bruijne}, J.~H.~J., {et~al.} 2016, \aap, 595, A1

\bibitem[{Gonz{\'a}lez-Casanova} \& {Lazarian}(2017)]{2017ApJ...835...41G}
{Gonz{\'a}lez-Casanova}, D.~F., \& {Lazarian}, A. 2017, \apj, 835, 41

\bibitem[{Goodman} {et~al.}(1998)]{1998ApJ...504..223G}
{Goodman}, A.~A., {Barranco}, J.~A., {Wilner}, D.~J., \& {Heyer}, M.~H. 1998, \apj, 504, 223

\bibitem[{Hacar} {et~al.}(2016)]{2016A&A...591A.104H}
{Hacar}, A., {Alves}, J., {Burkert}, A., \& {Goldsmith}, P. 2016, \aap, 591, A104

\bibitem[{Hern{\'a}ndez-Padilla} {et~al.}(2020)]{2020ApJ...901...11H}
{Hern{\'a}ndez-Padilla}, D., {Esquivel}, A., {Lazarian}, A., {et~al.} 2020, \apj, 901, 11

\bibitem[{Heyer} \& {Dame}(2015)]{2015ARA&A..53..583H}
{Heyer}, M., \& {Dame}, T.~M. 2015, \araa, 53, 583

\bibitem[{Heyer} \& {Brunt}(2004)]{2004ApJ...615L..45H}
{Heyer}, M.~H., \& {Brunt}, C.~M. 2004, \apjl, 615, L45

\bibitem[{Heyer} \& {Peter Schloerb}(1997)]{1997ApJ...475..173H}
{Heyer}, M.~H., \& {Peter Schloerb}, F. 1997, \apj, 475, 173

\bibitem[{Heyer} {et~al.}(2009)]{2009ApJ...699.1092H}
{Heyer}, M., {Krawczyk}, C., {Duval}, J., \& {Jackson}, J.~M. 2009, \apj, 699, 1092

\bibitem[{Huang} {et~al.}(2023)]{2023ApJ...949...46H}
{Huang}, B., {Wang}, K., {Girart}, J.~M., {et~al.} 2023, \apj, 949, 46

\bibitem[{Kainulainen} {et~al.}(2009)]{2009A&A...508L..35K}
{Kainulainen}, J., {Beuther}, H., {Henning}, T., \& {Plume}, R. 2009, \aap, 508, L35

\bibitem[{Kauffmann} {et~al.}(2008)]{2008A&A...487..993K}
{Kauffmann}, J., {Bertoldi}, F., {Bourke}, T.~L., {Evans}, N.~J., I., \& {Lee}, C.~W. 2008, \aap, 487, 993

\bibitem[{Kauffmann} {et~al.}(2013)]{2013ApJ...779..185K}
{Kauffmann}, J., {Pillai}, T., \& {Goldsmith}, P.~F. 2013, \apj, 779, 185

\bibitem[{Keto} \& {Myers}(1986)]{1986ApJ...304..466K}
{Keto}, E.~R., \& {Myers}, P.~C. 1986, \apj, 304, 466

\bibitem[{Klessen}(2000)]{2000ApJ...535..869K}
{Klessen}, R.~S. 2000, \apj, 535, 869

\bibitem[{Kolmogorov}(1941)]{1941DoSSR..30..301K}
{Kolmogorov}, A. 1941, Akademiia Nauk SSSR Doklady, 30, 301

\bibitem[{Kritsuk} {et~al.}(2013)]{2013MNRAS.436.3247K}
{Kritsuk}, A.~G., {Lee}, C.~T., \& {Norman}, M.~L. 2013, \mnras, 436, 3247

\bibitem[{Ladd} {et~al.}(1994)]{1994ApJ...433..117L}
{Ladd}, E.~F., {Myers}, P.~C., \& {Goodman}, A.~A. 1994, \apj, 433, 117

\bibitem[{Larson}(1981)]{1981MNRAS.194..809L}
{Larson}, R.~B. 1981, \mnras, 194, 809

\bibitem[{Leroy} {et~al.}(2016)]{2016ApJ...831...16L}
{Leroy}, A.~K., {Hughes}, A., {Schruba}, A., {et~al.} 2016, \apj, 831, 16

\bibitem[{Li} \& {Goldsmith}(2012)]{2012ApJ...756...12L}
{Li}, D., \& {Goldsmith}, P.~F. 2012, \apj, 756, 12

\bibitem[{Li}(2017)]{2017MNRAS.465..667L}
{Li}, G.-X. 2017, \mnras, 465, 667

\bibitem[{Li} {et~al.}(2022)]{2022ApJS..262...42L}
{Li}, J.~J., {Immer}, K., {Reid}, M.~J., {et~al.} 2022, \apjs, 262, 42

\bibitem[{Ma} {et~al.}(2022)]{2022ApJS..262...16M}
{Ma}, Y., {Wang}, H., {Zhang}, M., {et~al.} 2022, \apjs, 262, 16

\bibitem[{Mangum} \& {Shirley}(2015)]{2015PASP..127..266M}
{Mangum}, J.~G., \& {Shirley}, Y.~L. 2015, \pasp, 127, 266

\bibitem[{Mangum} \& {Shirley}(2016)]{2016PASP..128b9201M}
{Mangum}, J.~G., \& {Shirley}, Y.~L. 2016, {Corrigendum: How to Calculate Molecular Column Density}, Publications of the Astronomical Society of the Pacific, Volume 128, Issue 960, pp. 029201 (2016).

\bibitem[{Mei} {et~al.}(2024)]{2024A&A...685A..39M}
{Mei}, J., {Chen}, Z., {Jiang}, Z., {Zheng}, S., \& {Feng}, H. 2024, \aap, 685, A39

\bibitem[{Milam} {et~al.}(2005)]{2005ApJ...634.1126M}
{Milam}, S.~N., {Savage}, C., {Brewster}, M.~A., {Ziurys}, L.~M., \& {Wyckoff}, S. 2005, \apj, 634, 1126

\bibitem[{Miville-Desch{\^e}nes} {et~al.}(2017)]{2017ApJ...834...57M}
{Miville-Desch{\^e}nes}, M.-A., {Murray}, N., \& {Lee}, E.~J. 2017, \apj, 834, 57

\bibitem[{Peretto} {et~al.}(2023)]{2023MNRAS.525.2935P}
{Peretto}, N., {Rigby}, A.~J., {Louvet}, F., {et~al.} 2023, \mnras, 525, 2935

\bibitem[{Qian} {et~al.}(2015)]{2015ApJ...811...71Q}
{Qian}, L., {Li}, D., {Offner}, S., \& {Pan}, Z. 2015, \apj, 811, 71

\bibitem[{Reid} {et~al.}(2014)]{2014ApJ...783..130R}
{Reid}, M.~J., {Menten}, K.~M., {Brunthaler}, A., {et~al.} 2014, \apj, 783, 130

\bibitem[{Reid} {et~al.}(2019)]{2019ApJ...885..131R}
{Reid}, M.~J., {Menten}, K.~M., {Brunthaler}, A., {et~al.} 2019, \apj, 885, 131

\bibitem[{Rice} {et~al.}(2016)]{2016ApJ...822...52R}
{Rice}, T.~S., {Goodman}, A.~A., {Bergin}, E.~A., {Beaumont}, C., \& {Dame}, T.~M. 2016, \apj, 822, 52

\bibitem[{Riener} {et~al.}(2020{\natexlab{a}})]{2020A&A...633A..14R}
{Riener}, M., {Kainulainen}, J., {Beuther}, H., {et~al.} 2020{\natexlab{a}}, \aap, 633, A14

\bibitem[{Riener} {et~al.}(2020{\natexlab{b}})]{2020A&A...640A..72R}
{Riener}, M., {Kainulainen}, J., {Henshaw}, J.~D., \& {Beuther}, H. 2020{\natexlab{b}}, \aap, 640, A72

\bibitem[{Riener} {et~al.}(2019)]{2019A&A...628A..78R}
{Riener}, M., {Kainulainen}, J., {Henshaw}, J.~D., {et~al.} 2019, \aap, 628, A78

\bibitem[{Rosolowsky} {et~al.}(2021)]{2021MNRAS.502.1218R}
{Rosolowsky}, E., {Hughes}, A., {Leroy}, A.~K., {et~al.} 2021, \mnras, 502, 1218

\bibitem[{Sakai} {et~al.}(2022)]{2022PASJ...74..209S}
{Sakai}, N., {Nakanishi}, H., {Kurahara}, K., {et~al.} 2022, \pasj, 74, 209

\bibitem[{Scalo}(1984)]{1984ApJ...277..556S}
{Scalo}, J.~M. 1984, \apj, 277, 556

\bibitem[{Solomon} {et~al.}(1987)]{1987ApJ...319..730S}
{Solomon}, P.~M., {Rivolo}, A.~R., {Barrett}, J., \& {Yahil}, A. 1987, \apj, 319, 730

\bibitem[{Su} {et~al.}(2019)]{2019ApJS..240....9S}
{Su}, Y., {Yang}, J., {Zhang}, S., {et~al.} 2019, \apjs, 240, 9

\bibitem[{Sun} {et~al.}(2020)]{2020ApJ...901L...8S}
{Sun}, J., {Leroy}, A.~K., {Schinnerer}, E., {et~al.} 2020, \apjl, 901, L8

\bibitem[{Traficante} {et~al.}(2018)]{2018MNRAS.477.2220T}
{Traficante}, A., {Duarte-Cabral}, A., {Elia}, D., {et~al.} 2018, \mnras, 477, 2220

\bibitem[{Tritsis} {et~al.}(2022)]{2022MNRAS.514.3593T}
{Tritsis}, A., {Bouzelou}, F., {Skalidis}, R., {et~al.} 2022, \mnras, 514, 3593

\bibitem[{Tritsis} \& {Tassis}(2018)]{2018Sci...360..635T}
{Tritsis}, A., \& {Tassis}, K. 2018, Science, 360, 635

\bibitem[{V{\'a}zquez-Semadeni} {et~al.}(2019)]{2019MNRAS.490.3061V}
{V{\'a}zquez-Semadeni}, E., {Palau}, A., {Ballesteros-Paredes}, J., {G{\'o}mez}, G.~C., \& {Zamora-Avil{\'e}s}, M. 2019, \mnras, 490, 3061

\bibitem[{V{\'a}zquez-Semadeni} {et~al.}(2024)]{2024arXiv240810406V}
{V{\'a}zquez-Semadeni}, E., {Palau}, A., {G{\'o}mez}, G.~C., {et~al.} 2024, arXiv e-prints, arXiv:2408.10406

\bibitem[{VERA Collaboration} {et~al.}(2020)]{2020PASJ...72...50V}
{VERA Collaboration}, {Hirota}, T., {Nagayama}, T., {et~al.} 2020, \pasj, 72, 50

\bibitem[Virtanen {et~al.}(2020)]{2020SciPy-NMeth}
Virtanen, P., Gommers, R., Oliphant, T.~E., {et~al.} 2020, Nature Methods, 17, 261

\bibitem[{Wilson} {et~al.}(1970)]{1970ApJ...161L..43W}
{Wilson}, R.~W., {Jefferts}, K.~B., \& {Penzias}, A.~A. 1970, \apjl, 161, L43

\bibitem[{Xing} \& {Qiu}(2022)]{2022RAA....22g5006X}
{Xing}, Y., \& {Qiu}, K. 2022, Research in Astronomy and Astrophysics, 22, 075006

\bibitem[{Xu} {et~al.}(2021)]{2021ApJS..253....1X}
{Xu}, Y., {Bian}, S.~B., {Reid}, M.~J., {et~al.} 2021, \apjs, 253, 1

\bibitem[{Yan} {et~al.}(2021)]{2021ApJ...922....8Y}
{Yan}, Q.-Z., {Yang}, J., {Su}, Y., {et~al.} 2021, \apj, 922, 8

\bibitem[{Yan} {et~al.}(2022)]{2022AJ....164...55Y}
{Yan}, Q.-Z., {Yang}, J., {Su}, Y., {et~al.} 2022, \aj, 164, 55

\bibitem[{Zhang} {et~al.}(2024)]{2024AJ....167..220Z}
{Zhang}, S., {Su}, Y., {Chen}, X., {et~al.} 2024, \aj, 167, 220

\bibitem[{Zhao} {et~al.}(2022)]{2022ApJ...934...45Z}
{Zhao}, M., {Zhou}, J., {Hu}, Y., {et~al.} 2022, \apj, 934, 45

\bibitem[{Zhou} {et~al.}(2023)]{2023arXiv231201497Z}
{Zhou}, J.~W., {Dib}, S., {Wyrowski}, F., {et~al.} 2023, arXiv e-prints, arXiv:2312.01497

\bibitem[{Zhou} {et~al.}(2022)]{2022MNRAS.513..638Z}
{Zhou}, J.-X., {Li}, G.-X., \& {Chen}, B.-Q. 2022, \mnras, 513, 638

\bibitem[{Zhuang} {et~al.}(2024)]{2024ApJ...966..202Z}
{Zhuang}, Z., {Su}, Y., {Zhang}, S., {et~al.} 2024, \apj, 966, 202

\bibitem[{Zucker} {et~al.}(2021)]{2021ApJ...919...35Z}
{Zucker}, C., {Goodman}, A., {Alves}, J., {et~al.} 2021, \apj, 919, 35

\end{thebibliography}

\appendix

\section{Column Density Calculation} \label{ap:columndensity}

The column density for each line-of-sight can be calculated through
\begin{equation}
\begin{split}\label{eq:column_density}
    N_{\rm{tot}} = 2.48\times 10^{14} (T_{\rm{ex}}+0.88) & \exp \Big(\frac{E_u}{T_{\rm{ex}}}\Big) \Big[\exp \Big(\frac{E_u}{T_{\rm{ex}}}\Big) - 1 \Big]^{-1} \\ 
    & \times \frac{\int T_{\rm{MB}} dv~~(\rm{km~s^{-1}})}{f(J_{\nu}(T_{\rm{ex}}) - J_{\nu}(T_{\rm{bg}}))} \\
    & \times \frac{\tau}{1-\exp(-\tau)} ~\rm{cm^{-2}}
\end{split}
\end{equation}
which is derived under LTE assumption for \CO{}{18}{(J=1-0)} \citep{2015PASP..127..266M,2016PASP..128b9201M}
and has been corrected by the central optical depth
\begin{equation} \label{eq:tau}
    \tau = -\ln \left[1 - \frac{\T{peak}{}}{f(J_\nu(\T{ex}{})-J_\nu(\T{bg}{}))} \right].
\end{equation}
The excitation temperature \T{ex}{} can be derived from \T{peak}{12} and \CO{12}{}{~(J=1-0)} rest frequency $\nu$:
\begin{equation}
\T{ex}{} = \frac{h\nu}{k}\left[\ln \left(1 + \frac{h\nu/k}{\T{peak}{12} + J_{\nu}(\T{bg}{})}\right)\right]
\end{equation}
where $J_{\nu}(T)$ is the Rayleigh-Jeans equivalent temperature
\begin{equation}
J_{\nu}(T) \equiv \frac{h\nu/k}{\exp\left(\frac{h\nu}{kT}\right)-1},
\end{equation}
and \T{bg}{} is the background radiation temperature set to that of Cosmic Microwave Background (CMB) $\T{CMB}{} \simeq 2.73~\rm{K}$.

Eq. (\ref{eq:column_density}) can also be applied to \CO{13}{}{(J=1-0)} by adopting 
its upper level energy $E_{u} = 5.29\Kelvin$.
The derivation of the constant $2.48\times 10^{14}$ requires the dipole moment $\mu$ 
and rigid rotor rotation constant $B_0$ of the molecule. 
Substituting the corresponding $\mu$ and $B_0$ values for \CO{13}{}{} molecule\footnote{From \url{https://splatalogue.online/}} 
leads to the same constant in the precision of three significant digits. 
Therefore, the only difference in calculating the column density between the two emission lines is the upper level energy $E_u$.
Besides, $f$ is the beam filling factor that is always assumed to be identity in our cases.
\end{document}